\documentstyle[aps,prl]{revtex}  
\begin{document}
\flushbottom
\preprint{UCLA-NT-9802} 
\draft
 
\twocolumn[\hsize\textwidth\columnwidth\hsize  
\csname @twocolumnfalse\endcsname              

\title{\bf  Three-body decay of the $d^*$ dibaryon} 

\author { Chun Wa WONG }

\address{ Department of Physics and Astronomy, 
University of California, Los Angeles, CA 90095-1547  }

\date{\today}

\maketitle

\begin{abstract}
Under certain circumstances, a three-body decay width can 
be approximated by an integral involving a product of two 
off-shell two-body decay widths. This ``angle-average'' 
approximation is used to calculate the $\pi NN$ decay width of 
the $d^*(J^\pi=3^+, T=0)$ dibaryon in a simple $\Delta^2$ model 
for the most important Feynman 
diagrams describing pion emissions with baryon-baryon recoil and 
meson retardation. The decay width is found to be about 
0.006 (0.07, 0.5) MeV at the $d^*$ mass of 2065 (2100, 2150) MeV 
for input dynamics derived from the Full Bonn potential. The 
smallness of this width is qualitatively understood as the result 
of the three-body decay being ``third forbidden''. The concept of 
$\ell$ forbiddenness and the threshold behavior of a three-body 
decay are further studied in connection with the $\pi NN$ decay 
of the dibaryon $d'(J^\pi=0^-, T=0 \,{\rm or}\, 2)$ where the idea 
of unfavorness has to be introduced. The implications of these 
results are briefly discussed. 
\end{abstract}

\pacs{PACS number(s): }

 ]  

\narrowtext

\section{ Introduction } 

Almost all theoretical dibaryons with exotic quark structures have masses above the $\pi NN$ threshold \cite{Won98}. They can thus decay into 
$\pi NN$ states. From the viewpoint of dibaryon searches, the 
most promising candidates should have the narrowest widths, 
otherwise they do not stand out clearly above the background. 
However, their $\pi NN$ widths should not be too small, otherwise 
too few pions will be available to help with the identification. 
Hence a qualitative understanding of the $\pi NN$ decays of these dibaryons is of considerable interest when contemplating an 
experimental search for these dibaryons.

A particularly interesting dibaryon is the $d^*$, of 
quantum numbers $J^\pi T = 3^+0$. The interest comes from the possibility that its mass might be unusually low, thereby 
indicating an unusual structure or dynamics \cite{Gold89,Wang92,Gold95}. 
The purpose of this paper is to give rough estimates of its 
$\pi NN$ decay width when it is treated approximately as an S-wave 
bound didelta state of intrinsic spin $S=3$. 

The $\pi NN$ decay of this model $d^*$ cannot occur via pion 
emission from one of 
the constituent $\Delta$'s because the spectator baryon will 
remain a $\Delta$. Hence the baryons must interact at least 
once to turn the spectator into a final-state nucleon. This is 
true too for pion emission from a virtual meson which must be made 
to appear in the system. With one pion emission vertex and 
two interaction vertices for the spectator interaction (referred 
to below as a recoil), the leading Feynman diagrams are processes 
containing three vertices, as shown in Figs. 1 and 2. 

Of the processes shown, Fig. 1a-c are recoil diagrams describing 
$\pi$ emission by interacting baryons. Fig. 1d-f are retardation 
diagrams describing emission from a baryon when a virtual meson 
is in the air. Figs. 1g-i are some of the relativistic 
corrections coming from $N\bar N$ pairs. Finally Fig. 2 gives 
the contributions for emission from an exchanged meson itself.
We shall calculate the decay width for the most important of these 
processes with the help of an approximate angle-average formula  developed in this paper.

This paper is organized as follows: The general nature of the 
$\pi NN$ decay of the model $d^*$ is described in Sec. II for P- 
and S-wave pion emissions. The concept of $\ell$-forbiddenness 
for describing the threshold behavior of three-body decays is 
introduced in analogy to that for nuclear $\beta$ decays. Both 
P-wave and S-wave pion emissions from baryons described by  
Fig. 1 are found to 
be ``third-forbidden''. The processes shown in Fig. 2 for
pion emission from virtual mesons in the leading order 
are found to vanish for both P-wave and S-wave pion emissions. 

In Sec. III, an angle-average approximation 
for the width of a three-body decay of the type shown in Fig. 1 
is obtained in the form of a sum of integrals over the product of 
two sets of off-shell two-body decay amplitudes. 
Numerical results are given in Sec. IV where this angle-average approximation is applied to Figs. 1a-c, using baryon dynamics 
derived from the Full Bonn potential \cite{Mac87} and from 
on-shell $NN$ $t$-matrices constructed from experimental $NN$
phase shifts \cite{Fra85}. In Sec. V, retardation and pair 
contributions are included as well. 

Sec. VI gives a discussion of the threshold behaviors of 
$\pi NN$ decays, which are controlled but not solely 
determined by the $\ell$-forbiddenness of the decay. The decay of 
the $d' (J^\pi = 0^-, T=0 \,{\rm or}\, 2)$ dibaryon is studied to illustrate the complications caused by decays with abnormally 
large power dependences on the external momenta near threshold.
 
Finally, brief concluding remarks are made in Sec. VII, where 
the implications of our results for the quark-delocalization 
model of $d^*$ \cite{Gold89,Wang92,Gold95} are touched upon. In particular, the calculated decay width of 70 keV at $m^* = 2100$ 
MeV seems to imply that there might be too few decay pions at 
low dibaryon masses to help with particle identification.

\section{ General considerations } 

The dibaryon $d^*$ is an isoscalar, high-spin ($J^\pi=3^+$) 
state. When treated as a $\Delta^2$ bound state, its dominant 
spatial component is $^7S_3$. Fig. 1 gives the lowest-order 
Feynman diagrams for its decay into the $\pi NN$ final state by 
pion emission from a baryon. For P-wave pion emission, the $NN$ 
system in the $\pi NN$ final state has $ST=01$, the possible 
partial waves being $^1D_2$ and $^1G_4$. Hence the $\pi NN$ decay 
of $d^*$ through this dominant $^7S_3$ component requires three 
units of baryon spin change, i.e. a rank-3 intrinsic spin tensor. 

This rank-3 spin tensor must be combined into a 
rotational scalar as the perturbation responsible for the decay 
by scalar multiplication with a rank-3 spatial tensor made up 
of internal or external momenta. The Feynman diagrams shown 
in Fig. 1 describe some of the leading-order pion emission by a 
baryon in our simple model of $d^*$. The dominant process involves 
a P-wave pion emission that depends linearly on the momentum 
${\bf p}_3 = {\bf p}_\pi$ of the emitted pion. This means that 
the remaining $qq$ or $BB$ interaction must generate a rank-two 
spatial tensor to trigger the decay. This cannot be done with a 
rank-0 central interaction, or a rank-1 spin-orbit interaction. 
A rank-2 tensor-force interaction is needed. Furthermore, the 
final $NN$ state with the lowest orbital angular momentum that 
appears from the initial dominant spatial $^7S_3$-state of $d^*$ is 
$^1D_2$.

To the extent that $d^*$ is dominated by the $\Delta^2$ component
containing two $t=3/2$ constituents, the tensor interaction for 
recoil must convert at least one of these constituents into a 
$t=1/2$ nucleon. Thus the tensor force must be an isovector 
interaction that arises, for example, from the exchange 
of an isovector meson such as $\pi$ or $\rho$. This means that 
the operator in the $BB$ interaction responsible for the decay in 
our simple $\Delta^2$ model of $d^*$ is
$(\mbox{\boldmath $\sigma$}_i \times 
\mbox{\boldmath $\sigma$}_j)^{(2)} 
(\mbox{\boldmath $\tau$}_i . \mbox{\boldmath $\tau$}_j)$, where 
$i$ and $j$ refer to two separate baryons, or to quarks in 
these baryons. The operators $\mbox{\boldmath $\sigma$}$ and 
$\mbox{\boldmath $\tau$}$ are Pauli spin operators for quarks, and generalized spin operators for baryons \cite{Sug69}.

It is useful to enumerate the possible components of both the 
initial $d^*$ and the $B^2$ states appearing after 
P- or S-wave pion emission. We shall use a baryon notation with 
baryon components restricted to 
only $\Delta$ and $N$ because these baryons have the lowest masses 
and the same spatial $s^3$ quark structure. They can therefore be expected to mix strongly with one another. Under the circumstances, 
the available states are

\begin{eqnarray}
d^*(J^\pi = 3^+, T=0): && \Delta^2[^7(S, D, G, I)]; 
\nonumber \\ && N^2[^3(D,G)].
\nonumber \\
B^2(J^\pi = 2^+/4^+, T=1): && N^2[^1(D,G)]; \nonumber \\
&& \Delta N[^5(S, D, G, I)], [^3(D, G)]; \nonumber \\
&& \Delta^2 [^5(S, D, G, I)], [^1(D, G)].
\nonumber \\
B^2(J^\pi = 3^-, T=1): && N^2[^3F]; \nonumber \\
&& \Delta N[^5(P, F, H)], [^3(P, F)]; \nonumber \\
&& \Delta^2 [^7(P, F, H)], [^3(P, F)].
\label{eq-components}
\end{eqnarray}
The decay processes shown in Figs. 1a-c can then be described by 
certain decay equations. For P-wave emission, they are

\begin{eqnarray}
\Delta^2(^7S_3) & \rightarrow & N^2(^3D_3) \rightarrow N^2(^1D_2) + 
\pi(\ell_\pi =1); \nonumber \\
\Delta^2(^7S_3) & \rightarrow &\Delta N(^5S_2)+\pi(\ell_\pi = 1),
\nonumber \\ 
&& \Delta N(^5S_2) \rightarrow N^2(^1D_2); \nonumber \\
\Delta^2(^7S_3) & \rightarrow & \Delta^2(^5S_2) + \pi(\ell_\pi = 1),
\nonumber \\
&& \Delta^2(^5S_2) \rightarrow N^2(^1D_2)~.
\label{eq-Pwave}
\end{eqnarray}

In this paper, we are particularly interested in low-mass 
dibaryons decaying close to its $\pi NN$ threshold. Such 
near-threshold decays show certain general kinematical features 
reminiscent of nuclear $\beta$ decays, features that originate
from the dominance of centrifugal potential barriers on the 
outgoing decay products, and are controlled by their orbital 
angular momenta. 
The transition amplitude can be expanded in a Taylor series in 
these final-state momenta of which only two, 
say ${\bf p}_\pi$ of the pion and ${\bf p}_N$ of one of the 
nucleons, are independent. If $\ell_\pi$ is the pion orbital 
angular momentum in the dibaryon 
c.m. frame, and $\ell_N$ is the relative orbital angular momentum 
of the $NN$ system in the final state, the leading terms in the 
Taylor expansion are necessarily of the type
${\cal Y}_{\ell_\pi m_\pi}({\bf p}_\pi) 
{\cal Y}_{\ell_N m_N}({\bf p}_N)$, 
where ${\cal Y}_{\ell m}({\bf p})$ is a solid spherical harmonic.
The decay width near threshold is then proportional to 
$(p_\pi^{\ell\pi} p_N^{\ell_N})^2$ together with additional 
factors coming from the three-body phase space which we shall 
find in Sec. \ref{sec-Lforbid} to be $p_\pi^2 p_N^2$. We shall call 
this decay $\ell$-forbidden when $\ell = \ell_\pi + \ell_N$ is 
nonzero. With $\ell_\pi = 1$ and $\ell_N = 2$, the P-wave pion 
decays of Eq. (\ref{eq-Pwave}) are all third-forbidden.

This concept of $\ell$-forbiddenness is familiar not only from 
nuclear $\beta$ decays, but also from the simpler processes of 
two-body decays. For example, the decay of the $\Delta$ 
resonance involves P-wave pion emission, of $\ell_\pi = 1$. Hence 
the decay is $\ell = 1$ forbidden in our language. The two-body 
phase space gives an additional factor of $p_\pi$. The decay width 
is then roughly proportional to $p_\pi^3$. This is a well-known result.

The $\ell$-forbiddenness of the decay $d^* \rightarrow \pi NN$ is determined as follows: The pion has both orbital and total angular momenta $\ell_\pi$, and parity $\pi_\pi = (-1)^{\ell_\pi + 1}$. 
The initial state has quantum numbers $J_i^\pi = 3^+, T =0$. Hence 
the quantum numbers of the most favorable final $NN$ state is $T=1$, 
$\ell_N = \vert J_i - \ell_\pi \vert$, and $\pi_N = \pi_i \pi_\pi$.
From these selection rules, we can determine that the decay 
$d^* \rightarrow \pi NN$ is $\ell = J_i = 3$ forbidden for 
$\ell_\pi = 0-3$. For higher $\ell_\pi > J_i$, it is 
$\ell = 2\ell_\pi - J_i$ forbidden.

These selection rules for $\pi NN$ decays have been specified 
in terms of the relative coordinate of the final $NN$ state, i.e. 
a Jacobi coordinate. This is the most natural choice because wave-function antisymmetrization and final-state interactions can then
be described most conveniently. However, it is not 
the only one possible. Selection rules can also 
be specified in terms of the one of the final-state nucleon 
coordinates. This choice could occasionally be more convenient, for example when the nucleon is a real spectator. An example of this 
choice will be given in Sec. \ref{sec-Lforbid}.

Let us now return to the decay of $d^*$. For S-wave pion emissions, 
the final $NN$ system should have odd spatial parity, and 
therefore $ST=11$. Only one $N^2$ partial wave is then possible, 
namely $^3F_3$. Figs. 1a-c for S-wave pion emission from a baryon
are described by the decays

\begin{eqnarray}
\Delta^2(^7S_3) & \rightarrow & N^2(^3D_3) \rightarrow N^2(^3F_3) + 
\pi(\ell_\pi =0); \nonumber \\
\Delta^2(^7S_3) & \rightarrow & \Delta N(^5P_3) + \pi(\ell_\pi = 0),
\nonumber \\
&& \Delta N(^5P_3) \rightarrow N^2(^3F_3), \nonumber \\
\Delta^2(^7S_3) & \rightarrow & \Delta^2(^7P_3) + \pi(\ell_\pi = 0),
\nonumber \\
&& \Delta^2(^7P_3) \rightarrow N^2(^3F_3)~,
\label{eq-Swave}
\end{eqnarray}
respectively. Like P-wave pion-emission amplitudes, the S-wave amplitudes interfere among themselves, but they add only 
incoherently to the P-wave contributions to the decay width.

Now the basic S-wave pion emission vertex is known \cite{Eri88} 
to be of order $w/m$ relative to P-wave pion emission, where the 
baryon (or quark) energy transfer $w = E - E'$ is expected to be significantly smaller than the baryon mass $m$ for low-mass $d^*$. 
Given the fact that these S-wave decays are also 
$\ell = 3$ forbidden, we conclude that they can be neglected in comparison with P-wave emissions, at least for the rough estimates 
attempted in this paper.

Other types of S-wave pion emission are possible if the Feynman 
diagrams involve orbitally or radially excited baryons $\Delta^*$ 
or $N^*$ as well. These additional baryon configurations can be 
expected to be even less important because the $mBB^*$ coupling 
constants are much smaller than the meson coupling constants to 
$\Delta$ or $N$ because of the spatial quark excitations.

Among higher-order diagrams not included in Fig. 1 are those 
decribing final-state interactions between the outgoing nucleons.
They are not expected to be very important because the nucleons 
have large relative orbital angular momenta in our final states. 
Thus re-scattering effects in $d^* \rightarrow \pi NN$ decays 
could be quite different from those of threshold pion production 
in nuclear reactions \cite{Lee93,Han95}.

Let us turn next to Fig. 2 describing pion emission from  
virtual $\pi-\rho$ mesons. The operators involved in Fig. 2a 
have been derived in connection with the study of 
$\pi\pi\rho$-exchange three-nucleon forces \cite{Coe83}. 
Two pion-emission operators appear: The first is 
$(\mbox{\boldmath $\sigma$}_2.{\bf q}) 
(\mbox{\boldmath $\tau$}_2 \times 
\mbox{\boldmath $\tau$}_1).\mbox{\boldmath $\Phi$}$, 
where ${\bf q}$ is the momentum of the virtual pion and 
$\mbox{\boldmath $\Phi$}$ is its wave function. 
The spin-isospin operators 
appearing here could be those of quarks or of baryons 
(including transition spins) depending on the dynamical model 
used. For simplicity, however, we shall visualize them in this 
section as generalized baryon spin operators.

This pion-emission operator describes 
S-wave pion emission. For the two diagrams in Fig. 2 together, 
S-wave pion emission involves the combination
$(\mbox{\boldmath $\sigma$}_1 + \mbox{\boldmath $\sigma$}_2).
{\bf q} (\mbox{\boldmath $\tau$}_1 \times \mbox{\boldmath $\tau$}_2).\mbox{\boldmath $\Phi$}$. This is a rank-1 spin 
operator. Hence the direct decay of the major component $\Delta^2(^7S_3)$ of $d^*$ (i.e. Fig. 2) to the only available 
$N^2$ final state of $^3F_3$ is 
not possible. The process has to go through a minor component of 
either $d^*$ or the final $B^2$ state. A minor component 
of the final $B^2$ state is one which contains one or more 
$\Delta$'s that must be converted to nucleons before the decay 
is completed.

The second operator appearing in Fig. 2a is
proportional to $(\mbox{\boldmath $\sigma$}_2.{\bf q}) 
\mbox{\boldmath $\sigma$}_1.({\bf p}_3 \times {\bf q})
(\mbox{\boldmath $\tau$}_2 \times \mbox{\boldmath $\tau$}_1).\mbox{\boldmath $\Phi$}$, obtainable also from 
$\pi\pi\rho$-exchange three-nucleon forces. It gives rise to 
a pion emission operator of the form
$-2(\mbox{\boldmath $\sigma$}_1 \times 
\mbox{\boldmath $\sigma$}_2).
[{\bf q} \times ({\bf p}_3 \times {\bf q})]
(\mbox{\boldmath $\tau$}_1 \times \mbox{\boldmath $\tau$}_2).\mbox{\boldmath $\Phi$}$,
describing P-wave pion emission. The intrinsic spin operator 
appearing in it is also only rank 1. It too cannot connect the 
dominant $\Delta^2(^7S_3)$ component of $d^*$ to the available 
$N^2$ final states $^1D_2$ or $^1G_4$. Hence P-wave pion 
emision by a virtual meson is again possible only through a minor 
component in the initial or the final state. 

To summarize, to the lowest order in the 
strong-interaction vertices, the decay $d^* \rightarrow \pi NN$ 
involves the processes described by Fig. 1 involving pion 
emissions from baryons.

\section{ Angle-average approximation for three-body decay widths } 

Estimates of the width of a three-body decay of the type shown 
in Fig. 1 are greatly facilitated by using an angle-average approximation developed in this section.

We begin by noting that Fig. 1a can be interpreted as the 
off-shell decay 
$N \rightarrow \pi N$ of a small $NN(^3D_3)$ component of $d^*$ 
that has one spectator nucleon on its energy shell. This component 
is generated perturbatively by the recoil interaction from the 
$\Delta^2(^7S_3)$ state. For this reason, the energy denominator 
that appears is the bound-state expression

\begin{equation}
D_a = m^* - 2 E({\bf p}_2),
\label{eq-da}
\end{equation}
where $m^*$ is the $d^*$ mass and $E$ is a nucleon energy in 
the $d^*$ c.m. frame. The $t$-matrix appearing in the recoil 
interaction is an off-shell $t$-matrix between two baryons 
calculated at the initial-state energy $\sqrt{s} = m^*$. 

On the other hand, Figs. 1b and c involve pion emission into 
abnormal or minor $\Delta^2$ or $\Delta N$ components of 
the final $B^2$ channels. These abnormal components are 
off-shell with finite spatial extensions when the available 
energy is below their breakup threshold. 

Figs. 1b and c are also similar in that their energy 
denominators are scattering-state quantities of the type

\begin{equation}
D_b = E({\bf p}_1) + E({\bf p}_2) - E_\Delta(-{\bf p}) 
- E({\bf p}-{\bf p}_3),
\label{eq-db}
\end{equation}

\begin{equation}
D_c = E({\bf p}_1) + E({\bf p}_2) - E_\Delta(-{\bf p}) 
- E_\Delta({\bf p}-{\bf p}_3),
\label{eq-dc}
\end{equation}
where $E_\Delta$ is a $\Delta$ energy in the $d^*$ c.m. frame. 
Neither energy denominator vanishes below the appropriate 
threshold. The off-shell $BB$ $t$-matrix is that calculated at 
the energy of the final-state $NN$ pair. 

The $n$-body decay of an object of mass $m^*$ can be expressed conveniently by the Fermi golden rule \cite{PDG96}:

\begin{eqnarray}
\Gamma = (2\pi) \int \left( 1\over {\rm d}E\right)
\prod_{i=2}^n \left [{\rm d}^3{\bf p}\over (2\pi)^3\right]_i 
\langle \vert H'_{\rm fi} \vert ^2 \rangle_{\rm spin}
\nonumber \\ \times \delta (m^*-E){\rm d}E\, ,
\label{eq-Gamma}
\end{eqnarray}
where $H'$ is the effective perturbing Hamiltonian. The $n$-body 
density of states that appears can be expressed nominally in 
terms of a product of densities of states of fewer bodies. If the 
spin-average squared matrix element 
$\langle \vert H'_{\rm fi} \vert ^2 \rangle_{\rm spin}$
is also factorable into appropriate factors referring to fewer 
bodies in the problem, the integral could be simplified to make 
the physics more transparent.

To study this possibility for $d^*$, we first note that each 
Feynman diagram of Fig. 1 has the simple form of
$H_2D^{-1}H_3$, where $H_i$ is one of the interactions and $D$ 
is the energy denominator between them. The three-body phase 
space is unfortunately made complicated by the requirement of 
energy conservation which leaves behind the integrations over 

\begin{equation}
\left( 1\over {\rm d}E_1\right)
\prod_{i=2}^3 {\rm d}^3{\bf p}_i  
=  E_1 E_2 E_3 {\rm d}E_2 {\rm d}E_3 {\rm d}^2\Omega_3 
{\rm d}\phi_{23}~.
\label{eq-ps3}
\end{equation}
Here use has been made of the momentum-conserving relation

\begin{equation}
E_1^2 = p_2^2 + p_3^2 + 2p_2p_3 \cos \theta_{23} + m_1^2
\label{eq-e1}
\end{equation}
in the form

\begin{equation}
{{\rm d}\cos \theta_{23}\over {\rm d}E_1} = {E_1\over p_2 p_3} .
\label{eq-de1}
\end{equation}
Thus only one set of angle integrations, here ${\rm d}^2\Omega_3$,
is formally identical to that for the decay in free space.

What is left of the second set of angle integrations, 
namely d$\phi_{23}$, has been partly changed by the requirement 
of energy-momentum conservation for all three particles. As a 
result, kinematical quantities such as $E_1$ and $\theta_{23}$ 
are no longer independent of each other. Many of the angular 
functions in the integrand become quite complicated. The energy denominator $D$ too could be a function of some of these angle variables.

We are interested here in an angle-average approximation 
obtained by adding an extra angle integration 
$(1/2){\rm d}\cos \theta_{23}$ to Eq. (\ref {eq-ps3}) to restore 
the full angle integration of ${\rm d}^2\Omega_2$. The idea is to 
undo the angular correlation between the matrix elements of the interactions $H_1$ and $H_2$ (so that the the angle integrations 
$d^2 \hat {\bf p}_i$ can be done independently), but 
keep the proper three-body phase space of the Dalitz plot \cite{PDG96,Kal64}.

Before this can be done, another complication has to be taken care 
of. The integrand $\langle \vert (H')_{\rm fi} \vert ^2 
\rangle_{\rm spin}$ contains the propagator factor $D^{-2}$ 
which depends on the external hadron energies $E_i$ and certain 
internal baryon energies. The external hadron energies are 
angle-independent, but some of the internal baryon energies do 
depend on the integration angles. If such angle-dependent 
internal energies are replaced by suitable angle-averaged values, 
for example by the replacement

\begin{equation}
\vert {\bf p}-{\bf p}_3 \vert \rightarrow 
\sqrt{ \langle{\bf p}^2 \rangle + {\bf p}_3^2},
\end{equation}
the factor $D^{-2}$ simplifies to an angle-independent 
approximant $\langle D^{-2}\rangle$ that can be taken 
out of the angle integrations.
   
The remaining factor in the integrand is now

\begin{eqnarray}
\langle \vert (H_2 H_3)_{\rm fi} \vert ^2 
\rangle_{\rm spin} =  \prod_{i=2}^3 
\langle \vert (H_i)_{\rm fi} \vert ^2 \rangle_{\rm spin},
\label{eq-fac}
\end{eqnarray}
where the state labels $i, f$ in $(H_i)_{fi}$ are the initial- 
and final-state labels appropriate for the interaction $H_i$. 
The two  sets of angle integrations can now be done independently, leading to the ``off-shell'' two-body decay widths

\begin{eqnarray}
\Gamma_i(W_i) & = & {4\pi p_i^2{\rm d}p_i\over(2\pi)^2{\rm d}W_i} 
\langle \vert (H_i)_{\rm fi} \vert ^2 \rangle_{\rm spin, angle}~,
\label{eq-gam}
\end{eqnarray}
where

\begin{eqnarray}
\langle \vert (H_i)_{\rm fi} \vert ^2 \rangle_{\rm spin, angle}
= {1\over 4\pi} \int \langle \vert (H_i)_{\rm fi} \vert ^2 \rangle_{\rm spin}
d^2 \hat {\bf p}_i~.
\end{eqnarray}
These decay widths are in general off-shell because the energy in 
each final state differs in general from that in the initital state. 
The effective energy $W_i$ in the density of states remains to be 
chosen. 

Eq. (\ref{eq-Gamma}) has thus been simplified to

\begin{eqnarray}
\Gamma \approx {1\over 2\pi} & \int &
{\rm d}W_2 {\rm d}W_3 \left( {E_1\over 2p_2p_3} \right) 
\nonumber \\
&& \Gamma_2(W_2) \langle D^{-2} \rangle \Gamma_3(W_3)\,. 
\label{eq-gamgam}
\end{eqnarray}
This is the basic angle-average approximation used
in this paper. In this expression, the effective energies $W_i$ 
are arbitrary integration variables to be chosen for convenience 
since the integral is actually independent of their choices. 
Whatever the choice, the remaining integrations should be 
performed over an appropriate Dalitz area that reproduces what is 
left of the three-body phase space after the angle integrations. 

We choose $W_3 = E_3$ so that $\Gamma_3$ is actually the 
static limit [where $m_\pi \ll m$(nucleon)] of the physical width 
for pion emission. We use $W_2 = 2E_2$ in order to make the width 
$\Gamma_2$ for Fig. 1a an off-shell $d^* \rightarrow NN$ decay 
width with a two-body density of states defined for the $N^2$ 
system of total energy $2E_2$. 

The actual $\pi NN$ decay amplitude is of course a sum over contributions from several Feynman diagrams, though limited in 
this section for convenience in presentation to only Figs. 1a-c. 
To treat their interference, we shall need the complex 
transition amplitudes for the $BB$ recoil

\begin{equation}
F_i = \Gamma_i^{1/2} {\rm e}^{{\rm i}\phi_i}~.
\end{equation}
The final decay width will then be the sum of contributions

\begin{equation}
\Gamma = \sum_{\alpha, \beta} \Gamma(\alpha, \beta), 
\quad {\rm with}\:\alpha,\beta = a, b, {\rm or}\, c\, ,
\label{eq-gam3}
\end{equation}
where

\begin{eqnarray}
\Gamma(\alpha,\beta) 
\approx & {4\over 2\pi} & \int 
{\rm d}E_2 {\rm d}E_3 \left( {E_1\over p_2p_3} \right)
\nonumber \\ 
&& \times F_{2\alpha}^*F_{2\beta}
\langle (D_\alpha D_\beta)^{-1} \rangle
F_{3\alpha}^*F_{3\beta}~,
\label{eq-gamab}
\end{eqnarray}
if there were no additional complications in the angle and spin 
averages for the ``off-diagonal'' terms.
An extra factor of 4 appears because there are two $\Delta$'s in 
$d^*$. The diagonal terms $\Gamma(\alpha, \alpha)$ are positive, 
while the interference terms $\Gamma(\alpha, \beta \not= \alpha)$ 
could be negative.

Further development of this formula is possible under 
special circumstances. Matrix elements at corresponding vertices 
in different diagrams can be related to one another through the 
quark model \cite{Won98,Bro75}. For example, the $\pi qq$ 
vertex for pion emission from quark $i$ can be taken to be 
the NR expression

\begin{equation}
V_{\pi qq} = {f_{\pi qq}\over m_\pi} 
(\mbox{\boldmath $\sigma$}_i.{\bf p}_3) 
(\mbox{\boldmath $\tau$}_i.\mbox{\boldmath $\Phi$})~,
\label{eq-piqq}
\end{equation}
where $\mbox{\boldmath $\Phi$}$ is the pion wave function, and 
$f_{\pi qq}=(3/5)f_{\pi NN}$ is the coupling constant. The NR 
quark model can then be used to relate the $B \rightarrow \pi N$ 
decay widths that appear in these 
Born diagrams to just the basic width

\begin{eqnarray} 
\Gamma_{3b}(p_3) = \Gamma(\Delta \rightarrow N\pi)~,
\label{eq-G3b}
\end{eqnarray}
together with additional factors that represent changes in the 
reduced matrix elements of operators appearing in the 
interaction. Additional details will be given in the Appendix. 
Re-scattering effects for the emitted pion will not 
be included in the present study.

The $BB \rightarrow NN$ transition is more complicated. The 
transition matrix element $(H_2)_{fi}$ in Figs. 1b and c should 
be calculated in the $d^*$ rest frame, not in the final-state 
$N^2$ c.m. frame where the dynamics is usually specified. This 
means that

\begin{eqnarray} 
F_{2i}(W_2) = \delta_\rho {\cal F}_{2i}(p_2^*),
\end{eqnarray}
where

\begin{eqnarray} 
\delta_\rho = \sqrt{\rho_2\over \rho_2^*} = 
\sqrt{p_2E_2\over p_2^*E_2^*}~,
\end{eqnarray}
and ${\cal F}$ is an off-shell transition amplitude containing 
the two-body density of states calculated in the $N^2$ c.m. frame 
where the nucleon momentum is

\begin{eqnarray} 
p_2^* = \sqrt{{1\over 4}m_{12}^2 - m^2},
\end{eqnarray}
and

\begin{eqnarray} 
m_{12} = (E_1 + E_2)^2 - p_3^2
\label{m12}
\end{eqnarray}
is the invariant mass of the $N^2$ system.

We could still use the NR quark model to relate the different 
operator matrix elements of different diagrams so that the same 
basic transition amplitude ${\cal F}_{2a}$ appears in different diagrams, albeit evaluated at different energies and with 
different additional factors. Finally, spin averages must be 
performed term by term for Eq. (\ref{eq-gam3}), as discussed in 
the Appendix. In this way, we get the final result for Figs. 1a-c of

\begin{eqnarray}
\Gamma_{ac} 
& \approx & {25\over12\pi} \int 
{\rm d}E_2 {\rm d}E_3 {\left(E_1\over p_2p_3\right)}
\Gamma_{3b}  
\nonumber \\ && 
\times \left\vert {{\cal F}_{2a}(p_2)\over D_a} + 
\delta_\rho \left ( {2{\cal F}_{2a}(p_2^*)\over 3D_b} + 
{{\cal F}_{2a}(p_2^*)\over 3D_c} \right ) \right\vert^2~,
\label{eq-Fsum}
\end{eqnarray}
involving angle-averaged quantities.
This is the decay-width expression actually used in our calculations. 

We could have used empirical $mBB$ coupling constants \cite{Sug69} 
or some other model of $mBB$ dynamics \cite{Bro75} instead of the 
NR quark model to calculate different $BB$ recoil amplitudes. The 
result will differ somewhat from those shown in the equation. 
However, the differences are not expected to be very large from 
the viewpoint of the rough estimates attempted here. So we shall 
not consider these alternatives in this paper.

Equation ({\ref{eq-Fsum}) has a particularly simple structure if
the recoil amplitude ${\cal F}_{2a}$ is calculated in the Born approximation or evaluated at a common nucleon momentum $p_2$ 
instead of at two different values. On ignoring the small difference 
in the density of states in different $N^2$ frames as well, we find 

\begin{eqnarray}
\Gamma 
\approx &{25\over12\pi} & \int 
{\rm d}E_2 {\rm d}E_3 {\left(E_1\over p_2p_3\right)}
\Gamma_{3b} \Gamma_{2a} 
\nonumber \\ && 
\times \left\vert {1\over D_a} + {2\over 3D_b} + 
{1\over 3D_c} \right\vert^2\,~, 
\label{eq-gamsum}
\end{eqnarray}
where

\begin{eqnarray} 
\Gamma_{2a} = \Gamma(d^* \rightarrow N^2(^1D_2))~.
\label{eq-G2a}
\end{eqnarray}
This very rough formula for $\Gamma$ shows that the relative 
importance of the diagrams is approximately controlled by the 
energy denominators $D_i$. 

To estimate these energy denominators, we note that an external 
nucleon has the median energy of 

\begin{eqnarray}
E({\bf p}_2) \approx \mbox{$1\over 2$}(m_N + E_{{\rm max}}) = 961 
\,(988)\,{\rm MeV},
\label{eq-medE2}
\end{eqnarray}
where the numerical value is for $m^* = 2100\,(2200)$ MeV. For 
small $d^*$ masses, the internal baryon energies that appear are 
dominated by the internal momentum {\bf p} of $d^*$ of radius 
$r^*$ (usually taken to be 0.7 fm):

\begin{equation}
\mathclose{<}{\bf p}^2\mathclose{>} = {9\over 16r^{*2}} \approx 
(210 \,{\rm MeV}/c)^2,
\label{eq-psq}
\end{equation}

\begin{equation}
E({\bf p}) \approx \sqrt{m_N^2 + 
\mathclose{<}{\bf p}^2\mathclose{>}} =  962\,{\rm MeV},
\label{eq-eNp}
\end{equation}

\begin{equation}
E_\Delta({\bf p}) \approx \sqrt{m_\Delta^2 + 
\mathclose{<}{\bf p}^2\mathclose{>}} = 1250\, {\rm MeV},
\label{eq-eDp}
\end{equation}
and $E_i({\bf p}-{\bf p}_3) \approx E_i({\bf p})$.
These gives the rough estimates at $m^* = 2100\,(2200)$ MeV of 

\begin{eqnarray}
D_a \approx 178 \,(224)\,{\rm MeV};\qquad \qquad \nonumber \\
D_b \approx -290\,(-236)\,{\rm MeV},\:
D_c \approx -578\,(-524)\,{\rm MeV}.
\label{eq-Ds}
\end{eqnarray}

These approximate energy denominators can be used in Eq. 
(\ref{eq-gamsum}) to show that the decay widths from the three 
diagrams taken individually are roughly in the ratio

\begin{eqnarray}
\Gamma (a, a):\Gamma (b, b):\Gamma (c, c) \approx 
1:(-0.41)^2:(-0.10)^2
\end{eqnarray}
at 2100 MeV, and 

\begin{eqnarray}
\Gamma (a, a):\Gamma (b, b):\Gamma (c, c) \approx 
1:(-0.63)^2:(-0.14)^2
\end{eqnarray}
at 2200 MeV. Because the denominator of Fig. 1a has sign opposite 
to those in Figs. 1b and c, the amplitude from this dominant 
process interferes destructively with those from the latter 
diagrams, as shown explicitly in the last two equations. This destructive interference leads to a total decay width for Figs. 
1a-c that is roughly an order of magnitude smaller than 
$\Gamma_a$ from Fig. 1a alone.

\section{ Results for $d^* \rightarrow \pi NN$ } 

To obtain a $\pi NN$ decay width $\Gamma$, we need to integrate 
Eq. (\ref{eq-Fsum}) over the three-body phase space left over
after angle integrations. It seems 
useful to give a rough order-of-magnitude estimate of the 
dominant term $\Gamma_a$ from Fig. 1a as we go over some of 
the technical details involved in the integration.

The two-body width for P-wave pion emission is best expressed in 
terms of the experimental decay width 
$\Gamma_\Delta \equiv \Gamma(\Delta\rightarrow \pi N)$ of 120 MeV 
for a free $\Delta$ of mass $m_\Delta = 1232$ MeV:

\begin{equation}
\Gamma_{3b} = \Gamma_\Delta 
{\left(p_3m_\Delta\over k^*E_N^*\right)} 
{\left(p_3\over k^*\right)}^2 
{\rm e}^{\alpha(k^{*2}-p_3^2)},
\label{eq-gam3b}
\end{equation}
where the starred parameters are in the $\Delta$ c.m. frame:

\begin{equation}
k^* = 227 \:{\rm MeV},\quad E_N^* = 966 \:{\rm MeV}~. 
\label{eq-kstar}
\end{equation}
The Gaussian factor comes from a baryon formfactor calculated by 
using Gaussian wave functions in a NR quark model. It depends on a parameter

\begin{equation}
\alpha = r_p^2/6 = 1.54 \times 10^{-6}\:{\rm MeV}^{-2}. 
\label{eq-alpha}
\end{equation}
related to the proton radius $r_p$. The value of $r_p = 0.6$ fm 
used here is the one commonly used in many NR quark models, 
including \cite{Gold89}. It is not the experimental proton 
charge radius. Note that the pion momentum $p_3$ appearing in Eq.(\ref{eq-gam3b}) is that in the $d^*$ rest frame, and that the distortion of the outgoing pion wave function has been neglected.

Momentarily ignoring the Gaussian factor from the 
baryon formfactor, we find for $m^* = $ 2100 (2200) MeV 

\begin{eqnarray}
\Gamma_{3b\,{\rm max}} = 59\:(256) \:{\rm MeV},\:
\nonumber \\
\left \langle {\Gamma_{3b}\over p_3} \right\rangle  
\approx {\Gamma_{3b\,{\rm max}}\over 2.5p_{3\,{\rm max}} }= 
0.14\:(0.38)\:; 
\label{eq-avggam3b}
\end{eqnarray}
at the maximum value $p_{3\,{\rm max}} = $ 165 (270) MeV of 
$p_3$. The average shown in the last expression involving a 
denominator factor of 2.5 is obtained as follows: For 
nonrelativistic (extremely relativistic) kinematics, the phase 
space is roughly rectangular in $p_3^2$ ($p_3$). The average then 
has a denominator factor of 2 (3). For most of the pion energies 
studied here, the kinematics is neither nonrelativistic and 
extremely relativistic. So we take a denominator factor of 2.5.

The two-body width $\Gamma_{2a}(p_2)$ for the two-body decay 
$d^*\rightarrow NN$ is much less familiar, but it has recently 
been studied in \cite{Won98}. It is given by Eq. (38) of that 
paper, an expression that works for any final nucleon momentum 
$p_2$ in the rest frame of the final-state $N^2$ system. It can be 
used for the off-shell momenta appearing in our diagrams, provided 
that the reduced mass $\mu_f^*$ appearing in the equation is taken 
to be $E_2/2$ (where $E_2 = \sqrt{p_2^2 + m^2}$) and not $m^*/4$. 

Different dynamical inputs are possible. We use the full Bonn 
(FB) potential in the Born approximation, a choice referred to 
below as the FB model. Of course, the 
Born approximation of a static potential is rather crude because 
it contains no off-shell effect, but it represents a familiar 
starting point. The resulting width parameter $\Gamma_{2a}(p)/p^2$ 
is shown by a solid curve in Fig. 3 as a function of the nucleon momentum $p$ in the c.m. frame of the final $NN$ state.

Certain re-scattering effects not included in Fig. 1 could be 
included by using empirical $NN$ $t$-matrices constructed by 
Love and Franey (LF) \cite{Fra85} from experimental $NN$ phase 
shifts. These empirical $t$-matrices are of course on-shell $NN$ 
$t$-matrices, but here evaluated at the same momentum transfer $q$ 
as the needed off-shell $t$-matrices. So no dynamical off-shell 
effects can be included here. To my best knowledge, there is no 
simple way to extrapolate these empirical $t$-matrices off-shell 
without using a realistic potential model.

These LF on-shell $t$-matrices are actually energy dependent. 
In Fig. 1a, the $t$-matix involved is that for the initial 
dibaryon state of mass $m^*$. We therefore use the empirical 
$t$-matrix at the $N^2$ energy of $\sqrt{s} = m^*$. 

Now the $d^* \rightarrow NN$ decay in free space gives two 
nucleons each of momentum $p_{\rm max}$. The resulting decay width 
is represented by a solid circle in Fig. 3. In Fig. 1a, on the other 
hand, we need decay widths at the same $\sqrt{s}=m^*$ but with 
off-shell nucleon momentum $p < p_{\rm max}$. They can also be calculated for the same $t$-matrix input from the decay width 
formula of \cite{Won98} because this formula works for 
any nucleon momentum. The result is shown as a dashed curve in 
Fig. 3, one for each $\sqrt{s}$ whose value (in MeV) is also given
near the curve. These $t$-matrices are said to be calculated 
at the (initial-state) dibaryon energy (or DBE).

In contrast, the recoil interaction in Figs. 1b or 1c takes place 
after pion emission, i.e. at lower energies. It involves an 
off-shell $t$-matrix defined at the energy of the final $N^2$ 
state, as in the analgous processes in $pp$ bremstrahlung 
\cite{Bro68,Her92,Ede96}. In the present rough estimate, it is convenient to interpret it as an off-shell $BB \rightarrow NN$ 
decay process at an intermediate $B^2$-state energy 
$\sqrt{s} < m^*$. This means that one should use points from the 
dashed curves in Fig. 3 at smaller c.m. energies. The figure shows 
that such decay widths are smaller because the empirical 
$t$-matrix is found to decrease rapidly as $\sqrt{s}$ decreases. 

Such a strong energy dependence originates from the strong energy 
dependence of certain empirical $NN$ phase parameters such as the 
mixing parameter $\epsilon_1$ in the $N^2(^3S_1 - ^3D_1)$ states. 
The LF $t$-matrices we use were constructed in \cite{Fra85} from 
the energy-dependent phase solution SP84 of Arndt and 
collaborators \cite{Arn83}. It is well-known that the 
energy-dependent mixing parameter $\epsilon_1$ from SP84 already 
varies more smoothly and  more slowly with energy 
than the so-called single-energy solutions. Nevertheless, it still 
has an energy dependence much stronger than that of any 
common $NN$ potential such as the FB potential. This situation can 
be seen in Fig. 15-7 of \cite{Mac87}. In other words, the empirical 
$t$-matrices and the FB potential do not represent exactly the 
same dynamical input when it comes to this tricky isovector 
tensor-force part of the $NN$ interaction.

Given this complication, and the other problem of missing off-shell dynamics, we decide to use a much cruder approach for these 
empirical $t$-matrices. Instead of using them at different 
effective $B^2$ energies dictated by each diagram, we shall use 
two extreme prescriptions meant to provide rough upper and lower bounds. 

The first prescription is to use $t$-matrices of the same $NN$ 
energy $\sqrt{s} = m^*$ in all decay diagrams, i.e. using only 
points lying on one of the dashed curves in Fig. 3. These 
$t$-matrices are thus calculated at the initial-state dibaryon 
energy, a prescription referred to below as the LF-DBE model. 
One can see from Fig. 3 that the resulting decay widths are larger 
than those calculated at variable energies. The DBE results for 
$\Gamma_{\pi NN}$ can therefore be expected to be rough upper 
bounds. Fig. 3 shows that these DBE results are likely to be 
close to those calculated with the FB dynamics. 

The second prescription uses $t$-matrices calculated at the lowest 
possible energies in order to generate a rough lower bound. These energies will be taken to be $NN$ energies (NNE) after the recoil, 
giving a prescription to be called the LF-NNE model. The 
resulting $\Gamma_{2a}$ decay widths are the free-space widths 
reported in \cite{Won98} and represented by solid circles in Fig. 3. 
In other words, the solid circles are now treated as a function of 
the nucleon momentum $p_2$ and used in all diagrams. 
We can see from Fig. 3 that the $\Gamma_{\pi NN}$ that results will 
be much smaller than those for the LF-DBE or FB models. Since the 
LF-DBE model does not contain the effects of decreased effective 
$B^2$ energies for the recoil after pion emission while the LF-NNE 
model uses $NN$ energies that are far too low, the actual LF 
results can be expected to be somewhere between these two LF models.

Although we have not calculated the full off-shell $t$-matrix 
from the FB potential, it is nevertheless useful to point out some expected re-scattering effects in it. We shall concentrate on 
$\Gamma_{2a}$, where the $\Delta^2(^7S_3) \rightarrow N^2(^3D_3)$ 
matrix element of the isovector tensor force has the same sign as 
the $N^2(^3S_1) \rightarrow N^2(^3D_1)$ matrix element, according 
to Eqs. (\ref{A-rmeBB}) and (\ref{A-rmeNN}). The long-range part 
of this isovector tensor force is from pion exchange and is 
attractive in the $N^2(^3S_1 - ^3D_1)$ states. Hence re-scatterings 
can be expected to enhance its contribution, i.e. to increase the 
width in Fig. 3 for small momenta. On the other hand, the short-range 
part of the tensor force is dominated by rho exchange, and is 
repulsive. Re-scatterings will then reduce its value, leading to 
a decreased width in Fig. 3 at large momenta. These considerations 
suggest that with re-scatterings, the width from the FB model 
is likely to move towards that for the LF-DBE model. Even though
we do not know the quantitative extent of the change, we do 
expect the actual results to be much closer to the LF-DBE or FB 
values than the LF-NNE values. This is especially true if the 
mixing parameter $\epsilon_1$ given by the FB potential is more realistic than the empirical values obtained from phase-shift anaylsis.

To summarize, the three models shown in Fig. 3 seem to present an interesting range of dynamics for our rough estimate of the 
$\pi NN$ decay width. 

Let us now return to our rough estimate of $\Gamma_a$ using the 
FB model for $\Gamma_{2a}$.  We 
average the dimensionless ``reduced'' width $\Gamma_{2a}/p_2$ by 
itself over a rectangular distribution of $p_2^2$ (for 
approximately nonrelativistic kinematics) to get the rough estimate

\begin{eqnarray}
\left \langle {\Gamma_{2a}\over p_2} \right\rangle  
& \approx & {1\over p_{2{\rm max}}^2}\int_0^{p_{2{\rm max}}^2}
{\Gamma_{2a}(p)\over p} d(p^2) \nonumber \\
& = & 0.010\:(0.022) 
\label{eq-avggam2a}
\end{eqnarray}
for $m^* = $ 2100 (2200) MeV.

The energy denominator $D_a$ is next factored out of the integral 
by using the median energy $\langle E_1\rangle $ of $E_1$ given 
in Eq. (\ref{eq-medE2}). After replacing the factor $E_1$ in the 
integrand itself by its median value, we are finally left 
with an integration over the $E_2E_3$ (or $N\pi$) Dalitz area \cite{Kal64}

\begin{eqnarray}
A = \int_{\rm Dalitz} dE_2 dE_3.
\label{eq-DalitzA}
\end{eqnarray}
This area turns out to be 1390 (6800) MeV$^2$. Hence 

\begin{eqnarray}
\Gamma_a 
& \approx & {25\over 12\pi} A
\left \langle {\Gamma_{2a}\over p_2} \right\rangle  
{ \langle E_1\rangle \over \langle D_a\rangle^2 }
\left \langle {\Gamma_{3b}\over p_3} \right\rangle  
\nonumber \\
& \approx & 0.04\:(0.7) \:{\rm MeV},
\label{eq-gamaa}
\end{eqnarray}
for the FB potential.
As expected, the three-body decay width increases sharply 
with increasing $m^*$. It is also rather small, even before 
the destructive interference with the other amplitudes is included.

It is now a simple matter to return to Eq. (\ref{eq-Fsum}) 
and perform an honest integration over the Dalitz area without 
making the separate averages decribed previously in this 
section. In the energy denominators, we use the actual external
energies $E({\bf p}_1)$ and $E({\bf p}_2)$ of each point in phase 
space, the approximate internal energies for $E_i({\bf p})$ shown 
in Eqs. (\ref{eq-eNp}) and (\ref{eq-eDp}), and the approximation 

\begin{equation}
E_i({\bf p}-{\bf p}_3) \approx \sqrt{m_i^2 + 
\mathclose{<}{\bf p}^2\mathclose{>} + p_3^2},
\label{eq-eNpp3}
\end{equation}
that is the leading term of a Legendre expansion in 
${\bf p}.{\bf p}_3$. The calculated results from Fig. 1a alone are 
shown in Fig. 4 as a light solid curve for the FB model, as open diamonds for the LF-DBE model, and as a light dashed curve the 
LF-NNE model. The results for Figs. 1a-c are shown as a heavy 
dotted curve for the FB model, solid squares for the LF-DBE model, 
and a dot-dashed curve for the LF-NNE model.

We see that the results from all three models are in better 
agreement the higher the $d^*$ mass. The FB and LF-DBE results are 
quite close to each other over the entire mass range, with the 
LF-DBE widths a little larger close to the threshold, and a little
smaller as $m^*$ increases beyond 2100 MeV. The LF-NNE widths are 
about an order 
of magnitude smaller than those from the other two models below 
2100 MeV. Hence the FB widths seem to be quite reasonable, 
being comfortably within our rough upper and lower bounds at low 
$d^*$ masses.

Figure 4 also shows clearly how the destructive intereference 
among Figs. 1a-c reduces their resultant decay width significantly, 
as we have already noted previously.

\section{ Retardation and pair contributions } 

The remaining six diagrams shown in Fig. 1 come in two groups: 
Figs. 1d-f describe the retardation contributions due to pion 
emissions from a baryon when a virtual meson is in the air, while 
Fig. 1g-i describe pion emission via the creation of a $B\bar B$ pair.

The propagator for the retardation diagram $\alpha$ contains the 
energy-denominator factors $D_0^{-1}D_{\alpha}^{-1}$, where the 
initial-state energy denominator

\begin{equation}
D_0 = m^* - \omega - E({\bf p}_2) - E_\Delta ({\bf p})
\label{eq-D0}
\end{equation}
is the same in all three diagrams. Note that the sum 
of the two intermediate-state baryon energies that appears is 
only about 2210 (2240) MeV for $m^*$ = 2100 (2200) MeV. That is, 
this energy sum increases very slowly beyond the sum of their 
rest energies (at 2170 MeV) as $m^*$ increases. Hence for 
$m^* \approx 2200$ MeV, $D_0$ is close to $- \omega$, where 
$\omega = \sqrt{m_m^2 + q^2}$ is the meson energy.

The final-state energy denominator $D_\alpha$ depends somewhat on 
the diagram $\alpha$ with:

\begin{eqnarray}
D_d & = & E({\bf p}_2) -\omega - E_\Delta({\bf p}),
\nonumber \\
D_e & = & E({\bf p}_1) -\omega - E(-{\bf p}-{\bf p}_3),
\nonumber \\
D_f  & = & E({\bf p}_1) -\omega - E_\Delta(-{\bf p}-{\bf p}_3)~.
\label{eq-Df}
\end{eqnarray}
All angle-dependent energies will eventually be approximated by 
angle-averaged energies.

We turn next to the integration over 
one of the internal momenta, say the meson momentum ${\bf q}$. 
(The other internal momentum ${\bf p}$ is then fixed by 
momentum conservation.) This integration works out the same way 
it does for the two-body transition amplitude 
${\cal F}$ of Fig. 1a-c \cite{Won98}. The major difference is 
that the meson propagator $\omega^{-2}$ for Figs. 1a-c is now 
replaced by the propagator $1/\omega D_0D_\alpha$ for a 
retardation diagram $\alpha$. The appearance of an additional 
$\omega$-dependent energy denominator has the consequence that 
the longer-range $\pi$-exchange contribution becomes relatively 
more important than in Figs. 1a-c.

This change of propagator causes some complication in the 
calculation. Since the additional energy denominator depends 
on the meson mass, it can be included only if
the $\pi$- and $\rho$-exchange contributions can be separated. 
The transition amplitude ${\cal F}$ from Figs. 1a-c is of 
course separable into the form:

\begin{equation}
{\cal F} = {\cal F}_\pi + {\cal F}_\rho
\label{eq-sepF}
\end{equation}
when the virtual-meson exchange is treated literally in the 
Born approximation implied by the figures. By the same argument, 
the retardation diagrams 1d-f taken literally as Born diagrams 
give rise to transition amplitudes ${\cal R}$ of the same separable 
form

\begin{equation}
{\cal R} = {\cal R}_\pi + {\cal R}_\rho.
\end{equation}

The situation is more complicated when re-scattering effects are 
also included, for example, by using the LF $t$-matrices. While 
the amplitude ${\cal F}$ can be calculated without the separation 
of terms shown in Eq. (\ref{eq-sepF}), the calculation of  
${\cal R}$ is not possible without it. Fortunately, each LF 
$t$-matrix is made up of a sum of three or four terms with 
different effective meson masses, although the function used 
differs somewhat from that appearing in one-boson 
exchange potentials. Hence the calculation of ${\cal R}$ is still possible if one interprets these terms as the separate 
contributions from different mesons or groups of mesons. 

This prescription is admittedly far from ideal, but even a 
theoretically calculated $t$-matrix may not be separable into 
terms describing single-meson exchanges. The reason is simply that 
re-scattering means the inclusion of terms with two or more 
exchanged mesons. Their correct treatment can only be based on 
Feynman diagrams. In spite of this limitation, a calculation using 
the empirical $t$-matrix should still be very informative. The 
actual formula used to calculate ${\cal R}$ for LF $t$-matrices will 
be given in the Appendix.

One final problem in using the LF $t$-matrices has to be 
mentioned. The longest-range part of their tensor $t$-matrices is 
not one with the pion mass, but one with twice the mass. The 
authors found the constraint to match the one-pion range  
``too restrictive'' \cite{Fra85}, though they were successful in constraining the longest-range term of the real part of the 
central $t$-matrices to the one-pion value. This range problem 
is probably not serious in generating the amplitude ${\cal F}$, 
but possibly more serious in the amplitudes ${\cal R}$ for 
retardation diagrams. This is because these diagrams contain 
one more energy denominator dependent on the meson energy $\omega$. 
If the meson mass is too large in these energy denominators, the resulting decay amplitudes might be too small. It is 
hard to tell if this is indeed the case for a numerically fitted 
function as there might be compensating terms in the fit, but 
we shall watch out for this possibility in the calculated results.
 
We turn next to the pair diagrams of Figs. 1g-i. Since we need a 
rank-3 spin operator for the decay 
$d^*[\Delta^2(^7S_3)] \rightarrow \pi NN$, the 
pion-emission vertex has to be spin-dependent, while the 
virtual-meson exchange from or to the pair must involve a rank-2 
spin operator. The spin-dependent part of the pion emission vertex 
with the creation or annihilation of a $B\bar B$ pair turns out to 
be of order $p/m$ (ratio of baryon momentum to baryon mass) relative 
to the spin-dependent pion emission vertex from a baryon. The rank-2 spin part of the $B^2\bar B \rightarrow B$ interaction turns out to 
be also of order $p/m$ relative to the rank-2 spin operator in the
nonrelativistic $BB \rightarrow BB$ interaction for both $\pi$ and $\rho$ exchanges. Furthermore, the additional energy denominator 
for these pair diagrams is larger in absolute value than those in 
Figs. 1a-f by roughly twice the baryon mass. The consequence is 
that these pair contributions to the decay amplitude can be expected 
to be roughly two orders of magnitude smaller than those for the 
other diagrams of Fig. 1 in the mass region of interest in this 
paper. They can therefore be neglected.

Returning now to the retardation amplitudes ${\cal R}$,  we need 
to add them to Eq. (\ref{eq-Fsum}) and perform spin averages in 
the same way as for Figs. 1a-c. Again, it will be convenient 
to express all operator matrix elements for the retardation 
diagrams in terms of that appearing in Fig. 1d for which the 
transition amplitude is ${\cal R}_{2d}$, where the subscript 2 
refers to the ``spectator'' nucleon. The final expression for 
the $\pi NN$ width for Figs. 1a-f turns out to have 
the same structure as Eq. (\ref{eq-Fsum}):

\begin{eqnarray}
\Gamma_{af}
& \approx & {25\over12\pi} \int 
{\rm d}E_2 {\rm d}E_3 {\left(E_1\over p_2p_3\right)}
\Gamma_{3b}  
\nonumber \\ && 
\times \left\vert {{\cal B}(a,d) + 
\delta_\rho \left ( {2\over 3}{\cal B}(b,e) + 
{1\over 3}{\cal B}(c,f) \right ) }\right \vert^2~,
\label{eq-Bsum}
\end{eqnarray}
where

\begin{eqnarray}
{\cal B}(a,d) & \approx & {\cal R}_{2d}(d;p_2) + 
{{\cal F}_{2a}(p_2)\over D_a}, \nonumber \\
{\cal B}(b,e) & \approx & {\cal R}_{2d}(e;p_2^*) + 
{{\cal F}_{2a}(p_2*)\over D_b}, \nonumber \\
{\cal B}(c,f) & \approx & {\cal R}_{2d}(f;p_2^*) + 
{{\cal F}_{2a}(p_2^*)\over D_c}
\label{eq-B}
\end{eqnarray}
are amplitudes for pion emission from a baryon. The diagram labels 
$a-f$ are kept in the expressions as they define the energy 
denominators to be used in the amplitudes.

To simplify the calculation of the retardation amplitudes, we use 
in the energy denominators $D_e$ and $D_f$ the approximation

\begin{eqnarray}
E_i(-{\bf p}-{\bf p}_3) - E(-{\bf p}_2-{\bf p}_3) \approx 
\nonumber \\ E_i(-{\bf p}) - E(-{\bf p}_2)~,
\end{eqnarray}
where $i = \Delta$ or $N$ (i.e., blank). The mean external 
nucleon energy defined in Eq. (\ref{eq-medE2}) is used for 
$E(\pm {\bf p}_2)$, and the r.m.s. momentum 
$\langle {\bf p}^2 \rangle^{1/2}$ of the initial state is used 
in the internal energies $E_i({\bf p})$.

The resulting $\pi NN$ decay width $\Gamma_{af}$ is 
shown as a solid curve for the FB model, and as solid circles for 
the LF-DBE model. For comparison, the partial width $\Gamma_a$ for 
Fig. 1a alone is also shown, as a light solid curve for the FB model, and open diamonds for the LF-DBE model. All these widths have been calculated for a $d^*$ radius of 0.7 fm. The width $\Gamma_{af}$ for 
a $d^*$ radius of $r^* = 0.9$ (0.5) fm is also shown in the figure as 
a long-dashed (dashed) curve.

The main feature seen in Fig. 5 is that the retardation 
diagrams have overcome the reduction caused by the destructive interference among Figs. 1a-c to give a more normal result 
quite close to $\Gamma_a$ for Fig. 1a alone. It is interesting 
that $\Gamma_{af}$ is greater than $\Gamma_a$ for the FB model, 
but smaller for the LF-DBE model. The difference has a simple 
explanation. 

First of all, we should recall that the decay at the 
$BB \rightarrow NN$ step is controlled by a competition between 
the long-range attractive tensor force from $\pi$-exchange and 
the shorter-range repulsive tensor force from $\rho$-exchange \cite{Won98}. This cancellation makes the width increase less 
rapidly with increasing dibaryon mass. It also explains why the 
width is smaller for a dibaryon of smaller size, as seen in 
Fig. 5. 

Now the retardation diagrams contain an extra energy denominator 
that depends on the meson mass, a feature that favors the 
contribution from the virtual meson of smaller mass. Consequently,
the attractive $\pi$-exchange contribution is re-inforced in the 
FB model. In the LF models,
the one-pion term is absent in the isovector tensor force. The 
terms that remain correspond to larger meson masses giving larger 
energy denominator, and hence smaller retardation contributions.

If this picture is correct, the FB result would appear to be 
more sensible. However, a careful study of both theoretical 
$t$-matrices calculated from potential models and empirical 
$t$-matrices constructed from phase shifts will be needed to 
confirm this interpretation. We have not calculated the results 
for the TF-NNE model, but they are likely to be much smaller.

Our general conclusion is that while the energy dependence of 
the $t$-matrix is not insignificant, the best estimate we have 
now is based on the FB model. According to Fig. 5, it gives a 
value of $\Gamma_{af}$ for Figs. 1a-f that is about 0.006 
(0.07, 0.5) MeV at $m^*$ = 2065 (2100, 2150) MeV.

\section{ $\ell$ forbiddenness and threshold behaviors}
\label{sec-Lforbid} 

It is now useful to discuss the threshold behavior of the decay 
width. We shall begin with the Dalitz area $A(m^*)$ defined in 
Eq. (\ref{eq-DalitzA}). In the NR limit, it varies with momenta 
roughly as $(p_{2\,{\rm max}} p_{3\,{\rm max}})^2$. In reality, 
the dependence is only a little weaker than this, as the dotted 
curve for the ratio

\begin{equation}
C_A(m^*) = A(m^*)/(p_{2\,{\rm max}} p_{3\,{\rm max}})^2
\end{equation}
of $A$ to its threshold behavior shows in Fig. 6. 

Using the NR dependence for the Dalitz area, we find a threshold behavior for the 3-body decay of 

\begin{equation}
\Gamma(m^*) \approx {\rm const}\, p_{3\,{\rm max}}^{2\ell_\pi +2} 
p_{2\,{\rm max}}^{2\ell_N +2}.
\label{eq-threshold}
\end{equation}
This dependence could be approximated by just 
$p_{2\,{\rm max}}^{2\ell +4}$, 
where $\ell = \ell_\pi + \ell_N$ is the degree of forbiddenness 
of the decay, but we shall not make this additional 
approximation here. 

For $d^* \rightarrow \pi NN$ with P-wave pion emission, we have 
$\ell_\pi = 1$, $\ell_N = 2$, and a decay that is third forbidden. 
To see how dominant this threshold power law is, we show in Fig. 6 
the ratios

\begin{equation}
C_\alpha(m^*) = \Gamma_\alpha(m^*)
/(p_{3\,{\rm max}}^4 p_{2\,{\rm max}}^6),
\end{equation}
where $\alpha = a$ and $af$, as functions of the $d^*$ mass for 
both the partial width 
$\Gamma_{a}$ from Fig. 1a only (dashed curve) and the width $\Gamma_{af}$ from Figs. 1a-f (solid curve). Both widths are 
calculated with the FB model of interactions. 

These curves show that the threshold behavior is further 
modified by momentum dependences due to dynamics 
and form factors. In fact, both decay widths behave roughly 
as $p_{3\,{\rm max}}^4 p_{2\,{\rm max}}^5$ over the first half of 
the mass range shown in the figure, a small reduction in 
the momentum power being used to simulate the significant momentum dependences shown in the figure. 

These results are relatively simple because the initial state is a 
relative S-state. The situation is more complicated for a relative 
D-state. The solid spherical harmonic ${\cal Y}_{2m}({\bf p})$ 
that appears in the initial-state wave function would have given 
rise to an additional factor of

\begin{equation}
p^2 = ({\bf p}_2 + {\bf q})^2,
\label{eq-Dstate}
\end{equation}
where ${\bf q}$ is the momentum of the virtual meson involved in 
the $BB$ recoil. Since this internal momentum is eventually 
integrated over, what survives of the $q^2$ term has the same 
threshold behavior as those from the S-state component of $d^*$. 
The remaining terms in the equation have additional powers of 
the external momentum $p_2$ left over, leading to amplitudes with 
higher powers of $p_2$.

We have shown in Sec. II that the $d^*$ decay by S-wave 
pion emission is also third forbidden. This S-wave emission 
comes from the second, or S-wave, term in the basic $\pi qq$ 
vertex calculated from the usual $\pi qq$ Lagrangian with 
either pseudovector or pseudoscalar coupling: 

\begin{eqnarray}
V_{\pi qq} = {\rm const} 
[-\mbox{\boldmath $\sigma$}.( {\bf p}-{\bf p}' ) +
(w/4m)\mbox{\boldmath $\sigma$}.( {\bf p}'+{\bf p} )]
\label{eq-Vpinn}
\end{eqnarray}
in the NR and $w << m$ limit \cite{Eri88}:
Here {\bf p} ({\bf p}') is the quark momentum before (after)  
pion emission, the momentum of the emitted pion being 
${\bf p}_3 = {\bf p} - {\bf p}'$, and $w = E - E'$ 
is the baryon energy transfer. One can show 
that for $d^*$ decay, the use of the second term, the S-wave 
$\pi qq$ vertex, instead of the first, or P-wave, vertex results 
in the substitution of one power of the pion momentum $p_3$ by one 
power of the $NN$ relative momentum 
$\vert{\bf p}_1 - {\bf p}_2\vert$ in the final state. This 
substitution leaves the threshold power law unchanged.

However, it is not true that the threshold power law is always
determined completely by the degree of forbiddenness of the decay. 
Occasionally the expected threshold term is absent. Then the 
surviving leading terms are higher-order Taylor terms with a 
momentum power greater than 
expected. This situation will be described as being ``unfavored'' because an increase in the power implies a reduction in the decay 
width near threshold. So the general rule for the threshold power 
law is that the power is not smaller than that determined by the 
$\ell$-forbiddenness of the decay, but it could be bigger if the 
decay is unfavored.

To support the assertion that some decay amplitudes are 
unfavored, I now describe an example which is interesting in its 
own right. It is the $\pi NN$ decay of that most promising of 
low-mass dibaryons, namely $d'$ ($J^\pi = 0^-, T=$ 0 or 2), which 
was first proposed for a dibaryon interpretation of the structures 
seen in the pionic double charge exchange reaction 
$nn(\pi^+, \pi^-)pp$ on nuclear neutrons \cite{Bil92,Vor94,Bro96}.

The decay equations for S- and P-wave pion emissions are, respectively

\begin{eqnarray}
d'(0^-, T=0, 2) \rightarrow  N^2(^1S_0) + \pi(\ell_\pi =0). 
\end{eqnarray}
\begin{eqnarray}
d'(0^-, T=0, 2) \rightarrow  N^2(^3P_1) + \pi(\ell_\pi =1); 
\end{eqnarray}
This shows that S-wave pion emission is $\ell = 0$ allowed, 
while P-wave pion emission is $\ell = 2$ forbidden. This 
classification suggests that S-wave pion emission would normally dominate the decay near threshold.

A more detailed analysis shows that this is not always true.
To see this, let us start with the simple models of $d'$ used 
in \cite{Gar97}: a $\Delta N$ bound state in a relative P-state for 
the $T=2$ dibaryon, and an $N^*N$ relative-S bound state involving 
a P-state $N^*$ for the $T=0$ dibaryon. 

In the $\Delta N$ model of $d'(T=2)$, the off-shell $\Delta$ can 
decay into a nucleon while the spectator nucleon can come out without 
further interaction. The pion emission vertex can be written 
in one of the following two forms when specialized to the three-body final state

\begin{eqnarray}
V_{\pi N\Delta} = {\rm const} 
[\mbox{\boldmath $\sigma$}.{\bf p}_3 +
(w/2m)\mbox{\boldmath $\sigma$}.{\bf p}_{12}]~,
\label{eq-Jacobiform}
\end{eqnarray}
\begin{eqnarray}
V_{\pi N\Delta} = {\rm const} 
\left[ \left(1-{w\over 4m}\right)\mbox{\boldmath $\sigma$}.{\bf p}_3 -
{w \over 2m}\mbox{\boldmath $\sigma$}.{\bf p}_2 \right]~,
\label{eq-spform}
\end{eqnarray}
where

\begin{eqnarray}
{\bf p}_3 = {\bf p}_\pi, \quad
{\bf p}_{12} = {1\over 2} ({\bf p}_1 - {\bf p}_2) 
= - {\bf p}_2 - {1\over 2}{\bf p}_3
\end{eqnarray}
in the dibaryon rest frame, and where $\mbox{\boldmath $\sigma$}$ 
now stands for a transition spin operator \cite{Sug69}. We shall 
call these expressions the Jacobi form and the single-particle
form, respectively.

The spectator nucleon (of momentum ${\bf p}_2$) is a real 
spectator when final-state interactions are neglected. Its P-state
wave function is 

\begin{eqnarray}
\psi_{1 m} ({\bf p}_2) = f_1 (p_2) {\cal Y}_{1 m}({\bf p}_2),
\end{eqnarray}
where the radial function $f_1$ is finite at $p_2 = 0$.

Using the single-particle form for the pion emission vertex, we 
find that the first, or P-wave, vertex gives a decay amplitude 
proportional to 
${\cal Y}_{1 m}({\bf p}_2) \mbox{\boldmath $\sigma$}.{\bf p}_3$, 
while the S-wave vertex gives one proportional to 
${\cal Y}_{1 m}({\bf p}_2) \mbox{\boldmath $\sigma$}.{\bf p}_2$. 
One can recognize the presence of a P-wave pion in the first 
decay amplitude, and an S-wave pion in the second amplitude. 
However, both amplitudes depend on the external hadron momenta 
to the same second power. 

It is easy to see why the naively-expected threshold term for 
S-wave pion emission is missing. The S-wave vertex 
is actually a three-body interaction that is P-wave in the 
spectator momentum ${\bf p}_2$. Multiplication into the initial 
P-state wave function gives an extra factor of $p_2^2$. In other 
words, the normal S-wave result at threshold, assumed to be 
momentum-independent, is simply absent. What survives is a 
higher-order term. We shall call attention to this abnormal 
situation by calling the decay unfavored.  

An amusing complication in this decay of the $\Delta N$ model 
$d'(T=2)$ is worth pointing out. In terms of the Jacobi momentum 
${\bf p}_{12}$ of the final $N^2$ state, the initial P-state 
becomes a mixture of P- and S-states:

\begin{eqnarray}
{\cal Y}_{1 m}({\bf p}_2) 
& = & {\cal Y}_{1 m}(-{\bf p}_{12} - {1\over 2}{\bf p}_3) 
\nonumber \\
& = & - {\cal Y}_{1 m}({\bf p}_{12}) 
- {1\over 2}{\cal Y}_{1 m}({\bf p}_3)~.
\end{eqnarray}
This is caused by the recoil of the $NN$ system on pion emission. 
The second term on the right-hand side describes the process where 
the pion carries away the orbital angular momentum leaving the 
$N^2$ pair in a relative S-state.

The Jacobi description gives four amplitudes: Two are 
similar to those already described: a term proportional to 
${\cal Y}_{1 m}({\bf p}_{12})\mbox{\boldmath $\sigma$}.{\bf p}_3$
for P-wave pion emission, and a term proportional to
${\cal Y}_{1 m}({\bf p}_{12})\mbox{\boldmath $\sigma$}.{\bf p}_{12}$
for S-wave pion emission. However, two additional terms appear:
${\cal Y}_{1 m}({\bf p}_3)\mbox{\boldmath $\sigma$}.{\bf p}_{12}$ describing P-wave emission from the S-wave vertex, and finally
${\cal Y}_{1 m}({\bf p}_3)\mbox{\boldmath $\sigma$}.{\bf p}_3$ describing S- and D-wave pion emissions from the P-wave vertex. 
These unexpected terms have their orbital angular momentum changed 
by baryon recoil on pion emission. The D-wave term comes with an $NN(^1D_2)$ pair in the final state. It is 
$\ell =4$ forbidden, and is therefore quite unimportant.
These recoil-induced emissions are present in the single-particle 
form as well, but one has to know where to find them.

One can see from Eq. (\ref{eq-Jacobiform}) that the 
recoil-induced S-wave pion emission 
amplitude is not less important than the direct S-wave pion 
emission amplitude from the S-wave vertex, because 
unlike the latter it is not reduced by the additional factor 
$w/2m$. However, it generates the same unfavored S-wave 
threshold behavior of a quadratic dependence on the external 
momenta and not the normal behavior independent of external 
momenta. So none of the S-wave decay amplitudes in our simple 
model of $d'(T=2)$ is normal, and all are unfavored by two 
powers of the external momenta.

The above analysis is helpful because it tells us how to 
generate S-wave emission amplitudes that are normal and 
therefore dominant near threshold: The unwanted momentum can 
be prevented from appearing as an external momentum in an 
S-wave emission amplitude if it can be changed into an internal 
momentum that is eventually integrated over. This means that the 
P-wave excitation of the initial state should 
not be in the relative $BB$ coordinate, but in a quark 
coordinate internal to a baryon. In 
other words, amplitudes with normal S-wave threshold behavior 
can only come from those components of $d'(T=2)$ containing a 
P-wave excited $\Delta^*$ or $N^*$. 

The contributing term can readily be isolated by using a 
shell-model configuration in which a P-state quark in $B^*$ 
becomes an S-state quark in a nucleon after pion emission. The 
angular integrations over the initial-state quark momentum 
${\bf p}$ has the form

\begin{eqnarray}
\int d^2\hat {\bf p} {\cal Y}_{1 m}({\bf p}) V_{\pi qq} 
\exp [-b({\bf p} - {\bf p}_3)^2]~,
\end{eqnarray}
where the solid spherical harmonic comes from the initial P-state 
and the Gaussian comes from the final S-state wave function.
Using the single-particle form Eq. (\ref{eq-spform}) of the 
$\pi NN$ vertex with 
${\bf p}_2 = - {\bf p}$, we find that the P-wave vertex term 
contributes an S-wave pion emission amplitude that is unfavored 
by a momentum factor $p_3^2$. It is the S-wave vertex term that
gives rise to the sought-for amplitude that is independent of 
external momenta. Such an amplitude is more favorable than 
P-wave emissions near threshold by two degrees of forbiddenness. 
It can be expected to dominate the decay width near threshold. 

It is true, however, that the S-wave vertex is weaker than the 
P-wave vertex by a factor $w/2m$, and, depending on the 
dynamical model for $d'$, the $B^*B$ components of $d'(T=2)$ 
might themselves be weak. So the quantitative impact of this 
normal S-wave decay amplitude at the dibaryon mass of 2065 MeV is 
model-dependent.

Exactly the same situation holds for $d'(T=0)$ when treated as an 
S-wave $N^*N$ bound state, where $N^*$ is a P-state baryon of 
isospin 1/2. To the extent that this component might 
actually be a significant if not major part of $d'(T=0)$, 
the decay width of this dibaryon can be expected to be 
actually larger than that for $d'(T=2)$ near threshold.

The qualitative view on the decay of $d'$ implied by our 
forbiddenness classification and threshold behaviors both agrees 
with and differs from those obtained in recent studies of the 
$d'$ decay width in interesting ways. For the same models of $d'$, 
we agree with Garcilazo \cite{Gar97} that the $T=0$ decay width 
should be larger than the $T=2$ width, but both our widths are 
likely to be much smaller than Garcilazo's results. This is because 
for $d'(T=0)$, the normal allowed 
S-wave decay comes only from the weak S-wave $\pi qq$ vertex,
while for $d'(T=2)$ without any $B^*$ component, the  
$\ell = 2$ forbidden decay width should be relatively small near threshold.

We agree with Obukhovsky {\it et al.} \cite{Obu97} in finding 
that the P-wave $\pi NN$ vertex causes a recoil-induced decay into 
S-wave pion and $NN(^1S_0)$ final state, and that the resulting 
decay amplitude has an abnormal quadratic dependence on the pion momentum. However, another decay amplitude of comparable 
importance, namely P-wave pion emission caused by the P-wave 
$\pi NN$ vertex and leading to the $NN(^3P_1)$ final state has 
not been included in their calculation.

We agree with Samsonov and Schepkin \cite{Sam98} that another 
problem with the calculation of \cite{Obu97} is that the S-wave 
$\pi NN$ vertex has been left out. Its inclusion is important as 
it gives rise to the dominant allowed S-wave emission 
near threshold. Much is unclear, however. The energy transfer $w$
appearing in it is sometimes positive and sometime negative. It is 
not clear what it averages out to be. It is also not clear how much 
of the S-wave dominance remains by the time we get to the dibaryon 
mass of 2065 MeV.

\section{ Discussion and Conclusions } 

Just as in nuclear $\beta$ decay, a general three-body decay 
near threshold depends sensitively on the orbital angular 
momenta of the decay particles in the final state. A 
classification of the decay widths 
based on different degrees of $\ell$-forbiddenness then becomes 
useful. Much diversity exists within this broad classification, primarily because some decays are weaker than normal (or are 
unfavored) due to the presence of additional powers of external 
momenta in their decay amplitudes. 

According to this classification, the decays 

\begin{eqnarray}
d^*(J^\pi = 3^+, T=0) & \rightarrow & \pi(\ell_\pi=0)NN(^3F_3)
\nonumber \\ & {\rm and \,} & \pi(\ell_\pi=1)NN(^1D_2)
\end{eqnarray}
of the dibaryon $d^*$ are both normal ``third-forbidden'' decays. 
Even among such normal decays, considerable diversity remains in 
the decay widths due to differences in wave functions and in dynamics. 

In this paper, we have used an angle-average approximation to 
make quick estimates of the $\pi NN$ decay width of a $d^*$ 
described as a simple $\Delta^2$ bound state. We find a 
leading-order decay width of only 70 keV at the 
$d^*$ mass $m^*$ of 2100 MeV. The calculated width decreases 
rapidly as $m^*$ decreases towards the threshold at 2020 MeV, 
being only about 6 keV at $m^*$ = 2065 MeV.

The accuracy of these angle-averaged results must be confirmed in 
the future by more detailed calculation. Such a calculation is 
conceptually simple, but a little tedious in execution as it 
involves additional angle integrations. The uncertainties caused 
by angle averaging are likely to be much smaller than the 
uncertainties caused by uncertainties in the wave functions and in 
the strong-interaction dynamics. 

Although the $d^*$ model used here is crude, the 
calculated results can be used to estimate the 
decay widths for other models of the wave function. 
For example, the $d^*$ wavefunction is much more complicated 
in the quark-delocalization and color-screening 
model of \cite{Gold89,Wang92,Gold95}. It has been estimated  \cite{Won98} that quark delocalization would decrease the present 
result by a factor of 0.4, while only 1/5 of the six-quark 
wavefunction is in the $\Delta^2$ configuration. The remainining 
4/5 of the $d^*$ state is in hidden-color configurations which 
will contribute much less, perhaps only 1/5 or less of the 
contribution of $\Delta^2$. 

The $d^*\rightarrow NN$ width used 
here might also have been overestimated by the meson-exchange 
model used in \cite{Won98}. This would be the case if the interior 
of the $d^*$ dibaryon is a perturbative vacuum where the 
exchanged mesons do not exist as such and where the dynamics is 
much weaker. Assuming that the hidden-color contribution is 
cancelled by the reduction caused by the perturbative vacuum, we 
are left with a total reduction, for the quark-delocalization and 
color-screening model of $d^*$, of about an order of magnitude 
from the results reported here for our simple model of $d^*$ as $\Delta^2$.

There are other wave-function uncertainties such as the presence 
of the $\Delta^2(^7D_3)$ component, which could be significant or negligible depending on the dynamics assumed to operate at short distances.
 
We have relied on the familiar meson-exchange model of nuclear 
forces to generate the dynamics needed in the calculation. Unfortunately, the calculated decay width involves the short-range 
part of the isovector tensor force where there is much  
cancellation between the $\pi$- and $\rho$-exchange contributions. 
Our knowledge of this particular combination of isovector tensor 
forces, and of short-distance dynamics in general, 
seems to be relatively poor.

In spite of these uncertainties, our results seem to suggest that 
the $\pi NN$ decay width of $d^*$ might be too small at low $d^*$ 
masses to make it easy to use the few decay pions for particle identification.

\acknowledgements

I would like to thank Stan Yen, Terry Goldman, Fan Wang and Gary 
Love for many helpful discussions and correspondence.

\appendix
\section{Spin averages and two-body decay amplitudes}

The three-body decay $d^*\rightarrow \pi NN$ involves two groups 
of two-body reduced matrix elements (RME) -- one for $\pi$ emission 
from a baryon and one for the two-baryon transition 
$BB\rightarrow NN$, where a baryon could be $\Delta$ or $N$. 
When brought down to the quark level, the RME's needed for $\pi$ emission from different baryons are related by the 
nonrelativistic (NR) quark model as follows:

\begin{eqnarray}
(\Delta \Vert \mbox{\boldmath $\sigma$}_1 \mbox{\boldmath $\tau$}_1
\Vert N) = {8\over 3}\sqrt{2},\qquad \qquad \nonumber \\
(\Delta \Vert \mbox{\boldmath $\sigma$}_1 \mbox{\boldmath $\tau$}_1
\Vert \Delta) = {20\over 3},\: \:
(N \Vert \mbox{\boldmath $\sigma$}_1 \mbox{\boldmath $\tau$}_1
\Vert N) = {10\over 3}~.
\end{eqnarray}
where quark 1 is in the baryon. 

The operator responsible for the $BB\rightarrow NN$ transitions 
needed in Figs. 1a-c is

\begin{eqnarray}
T_{\rm BB} = - (\mbox{\boldmath $\sigma$}_1 \times 
\mbox{\boldmath $\sigma$}_4)^{(2)} 
(\mbox{\boldmath $\tau$}_1 . \mbox{\boldmath $\tau$}_4)~,
\end{eqnarray}
where quarks 1 and 4 are in two different baryons. The following 
RME's are needed in our calculations:

\begin{eqnarray}
(\Delta^2, ST = 30 \Vert T_{\rm BB} \Vert N^2, 10) & = &
{16\over 9}\sqrt{{7\over 2}}, \nonumber \\
(N\Delta, ST = 21 \Vert T_{\rm BB} \Vert N^2, 01) & = &
-{40\over 27}\sqrt{{5}},\nonumber \\
(\Delta^2, ST = 21 \Vert T_{\rm BB} \Vert N^2, 01) & = &
{40\over 27}\sqrt{2}~.
\label{A-rmeBB}
\end{eqnarray}
For comparison, we also give the RME between $N^2$ states:

\begin{eqnarray}
(N^2, ST = 10 \Vert T_{\rm BB} \Vert N^2, 10)  = 
{50\over 27}\sqrt{5}.
\label{A-rmeNN}
\end{eqnarray}

The spin averages of the product of operator matrix elements 
appearing in the contributions to the decay width shown in 
Eq.(\ref{eq-fac}) are relatively easy to calculate directly for 
the $d^*$ S-state. For Fig. 1a alone, we have

\begin{eqnarray}
&& {1\over 2S+1} \sum_{M,M_{\rm T}} \sum_{\mu,m,n} 
\vert \langle i \vert 
(\mbox{\boldmath $\sigma$}_1 \times 
\mbox{\boldmath $\sigma$}_4)_\mu^{(2)} 
(\mbox{\boldmath $\tau$}_1 . \mbox{\boldmath $\tau$}_4)
\sigma_m \tau_n 
\vert f \rangle\vert^2 \nonumber \\
& = & {25\over 24} (\Delta^2, ST = 30 \Vert T_{\rm BB} \Vert N^2, 10)
(\Delta \Vert \mbox{\boldmath $\sigma$}_1 \mbox{\boldmath $\tau$}_1
\Vert N)~,
\end{eqnarray}
where 

\begin{eqnarray}
\vert i \rangle & = & \vert d^*; S=3M, T=00 \rangle,\nonumber \\
\vert f \rangle & = & \vert (\pi NN) 00, 1M_{\rm T} \rangle~.
\end{eqnarray}
The quark label has been left out of the quark operators 
$\sigma_m \tau_n$ responsible for the pion emission. 
We have also chosen to express this and other spin averages in 
terms of the RME's for $\Delta^2\rightarrow NN$ and 
$\Delta\rightarrow \pi N$ for convenience in mutual comparisons. 
The spin averages for the other diagrams and for interference 
terms can be calculated in the same way.

The pion momentum is the same for all Feynman diagrams, but 
the dependence on the nucleon momenta is more complicated. 
We shall assume that the $BB\rightarrow NN$ interaction 
should be evaluated in the c.m. frame of the baryons in Figs. 
1b, c, e and f, and not that of $d^*$. This means, for example, 
that the nucleon momentum appearing in the corresponding decay amplitudes for Figs. 1b and c, denoted ${\cal F}_{2a}$ in the 
text, is different from that for Fig. 1a, as explained more fully 
in Sect. II. This refinement is included in the final expression 
(\ref{eq-Bsum}), which also gives the results of all spin averages.

Two two-body decay widths (or amplitudes) appear in this equation. 
The one for the decay $\Delta \rightarrow \pi N$ is obtained from 
the free-space decay width $\Gamma_\Delta$ based on a P-wave 
$\pi qq$ vertex for pion emission from quarks. This means that we 
have neglected a certain well-known S-wave vertex term which when expressed in baryon coordinates is proportional to the mean 
momentum of the emitting baryon. This S-wave emission vanishes 
in free space because the appropriate frame to be used is the one 
where the mean momentum vanishes, i.e. the Breit frame. It does 
not vanish in a many-baryon system where the mean baryon momentum 
is nonzero. We have argued in Sec. II, however, that for $d^*$ 
decay this S-wave vertex is likely to be much less important than 
the more familiar P-wave vertex. It may therefore be neglected in 
the qualitative estimates attempted in this paper.

Two different dynamical inputs for the $BB\rightarrow NN$ 
transition amplitudes ${\cal F}_{2a}$ are used in Eq. 
(\ref {eq-Bsum}). It depends on the nucleon momentum $p_2^*$ in 
the $B^2$ rest frame of invariant mass $m_{12}$:

\begin{eqnarray}
{\cal F}_{2a}(p_2^*) & = & \left( \sqrt{p_2^*\mu^*}\over 2\pi \right) 
\sqrt{{8\over 1 5}} \left( {12\over 5} \right)^2 
\left( 1\over \pi\beta^{^*2} \right)^{3/4} 
\nonumber \\ 
&& \times e^{-\kappa^2/2}I(\kappa)
\left( \beta^{^*5}\over \kappa^3 \right)~,
\label{eq-calF}
\end{eqnarray}
where 

\begin{eqnarray}
\mu^* = m_{12}/4, \quad \kappa = p_2^*/\beta^*,\quad 
\beta^* = \sqrt{3/8}\,r^*,
\end{eqnarray}
$r^* $ is the $d^*$ radius, usually taken to be 0.7 fm, and 
$I(\kappa)$ is an integral \cite{Won98} over the momentum 
$q = \beta^* Q$ of the virtual meson responsible for the $BB$ recoil:

\begin{eqnarray}
I(\kappa) = {\rm e}^{-\kappa^2/2} & \int & {\rm e}^{-Q^2/2}
[j_2(i\kappa Q)(\kappa Q)^3] \nonumber \\
&& t_{\rm ivt}(\beta^*Q)Q {\rm d}Q\,.
\end{eqnarray}
The function $t_{\rm ivt}(q)$ is the radial part of the isovector 
tensor force appearing in both meson-$qq$ and meson-$BB$ interactions. 

For the Full-Bonn (FB) potential \cite{Mac87} in the Born 
approximation, we use
 
\begin{eqnarray}
t_{ivt}(q) \approx v_{ivt}(q) = \sum_i v_i(q),
\end{eqnarray}
where

\begin{eqnarray}
v_i(q) & = & 4\pi {S_i\alpha_i\over 4m_i^2} {1\over q^2 + m_i^2}
\left ( \Lambda_i^2 - m_i^2\over \Lambda_i^2 + q^2 \right )^2.
\end{eqnarray}
The sum is taken over the two virtual mesons $i = \pi, \rho$, 
with signatures $S_\pi = -1, S_\rho = 1$ and strengths

\begin{eqnarray}
\alpha_\pi & = & {g_\pi\over 4\pi} = 14.4, \nonumber \\
 \alpha_\rho & = &{g_v^2\over 4\pi}\left (1 + {f_v\over g_v} \right )^2
= 0.84 (1+6.1)^2~,
\end{eqnarray}
respectively. The other parameters are those shown on p. 37 of \cite{Mac87}.

For the Love-Franey $t$-matrix, we use Eq. (15c) of \cite{Lov81} 

\begin{eqnarray}
t_{ivt}(q) & = & \sum_i t_i(q), 
\end{eqnarray}
where
\begin{eqnarray}
t_i(q) & = & 32\pi {V_i^T q^2 R_i^7\over [1+(qR_i)^2]^3}.
\end{eqnarray}
This is a sum of 3--4 terms of different ranges $R_i$ 
corresponding to virtual mesons of different masses 
$m_i = \hbar c/R_i$. The parameters are given in Table I of \cite{Fra85}.

In the approximation described in Sec. V, two of the retardation amplitudes ${\cal R}$ involve the same function:

\begin{eqnarray}
{\cal R}_{2d}(f; p_2^*) = {\cal R}_{2d}(d; p_2^*)~.
\end{eqnarray}
Only two distinct functions are then needed. Both are generated 
from Eq. ({\ref{eq-calF}), but by using a certain retardation 
analogs $r_{ivt}$ of $t_{ivt}$. For the FB potential,
they are

\begin{eqnarray}
r_{ivt}(\alpha; q) & = & \sum_i {v_i(q)\omega_i\over
D_{0i}D_{\alpha i}},
\label{eq-Arivt}
\end{eqnarray}
where the argument $\alpha = d$ ($e$) denotes Fig. 1d (1e), 
$\omega_i = \sqrt{m_i^2 + q^2}$, and 

\begin{eqnarray}
D_{0i} & \approx & m^* - \omega_i - \langle E_2 \rangle 
- \langle E_\Delta \rangle, \nonumber \\
D_{\alpha i} & \approx & \langle E_2 \rangle - \omega_i 
- \langle E_B \rangle. 
\end{eqnarray}
The baryon energy $E_B$ is $E_\Delta$ for $\alpha = d$, 
and the nucleon energy $E$ for $\alpha = e$. The averages 
appearing here are those described in Sec. V.

For the LF $t$-matrices, the Born term $v_i$ in Eq. 
(\ref{eq-Arivt}) should be replaced by the $t$-matrix term $t_i$.

\centerline{\bf Figure Captions}

\begin{figure}
\caption{ Some leading-order Feynman diagrams for the decay
$d^* \rightarrow \pi NN$ with pion emission from a baryon. }
\end{figure}

\begin{figure}
\caption{ Some leading-order Feynman diagrams for the decay
$d^* \rightarrow \pi NN$ with pion emission from virtual mesons. }
\end{figure}

\begin{figure}
\caption{ The $d^* \rightarrow NN$ decay width parameter 
$\Gamma_{2a}(p)/p^2$ as a function of the off-shell nucleon 
momentum $p$ in the $NN$ c.m. frame for the Full Bonn (FB) 
and the Love-Franey (LF) models of dynamics. Each LF curve is 
labeled by the c.m. energy of the $NN$ system whose 
empirical free-space $t$-matrix is used in the calculation. The 
on-shell decay width at each energy is given by a solid circle 
in the figure. }
\end{figure}

\begin{figure}
\caption{ The widths $\Gamma_a$ from Fig. 1a alone and 
$\Gamma_{ac}$ from Figs. 1a-c for the decay 
$d^* \rightarrow \pi NN$ for different dynamical inputs: the Full 
Bonn potential (FB), Love-Franey $t$-matrices calculated at the 
dibaryon energy (LF-DBE) and at the $NN$ energy (LF-NNE). }
\end{figure}

\begin{figure}
\caption{ Ratios of the Dalitz area A, the decay width $\Gamma_a$ 
from Fig. 1a, and the decay width $\Gamma_{af}$ from Figs. 1a-f to 
the expected threshold power law for the $d^* \rightarrow \pi NN$ 
decay as functions of the dibaryon mass $m^*$. Each ratio is given
in units of the value at $m^*$ = 2034 MeV. }
\end{figure}

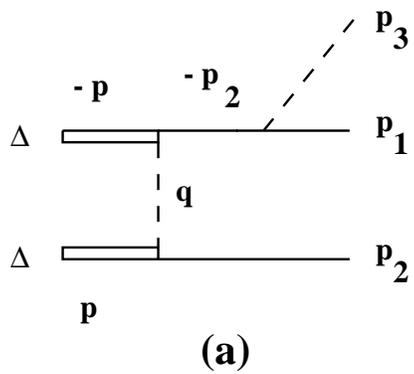
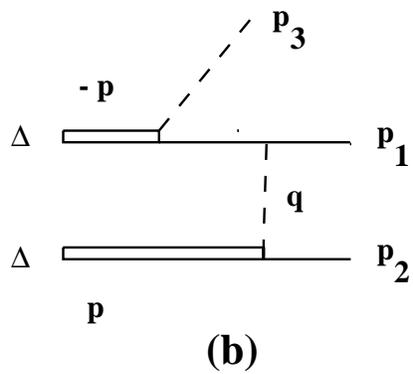
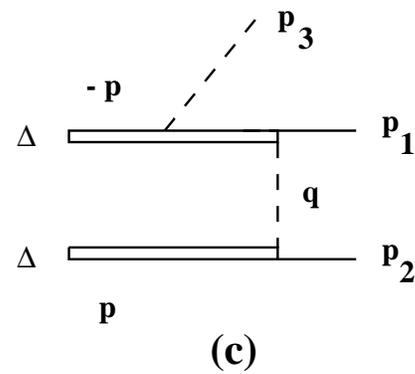
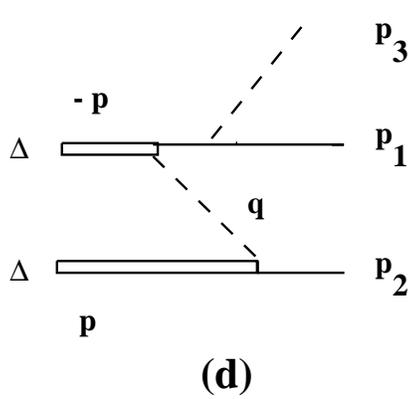
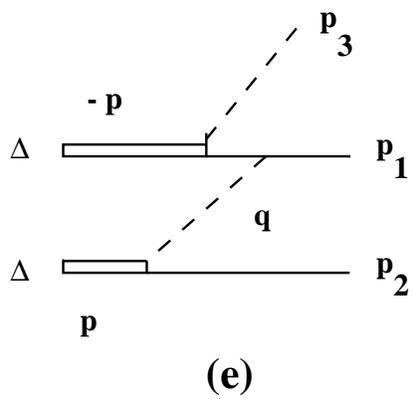
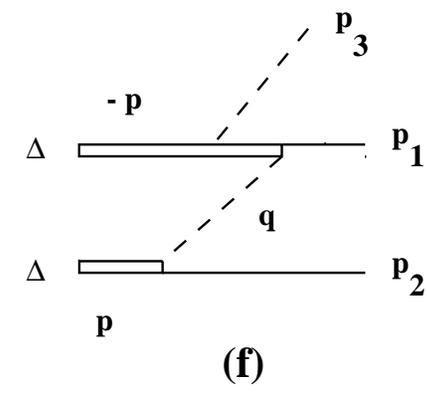
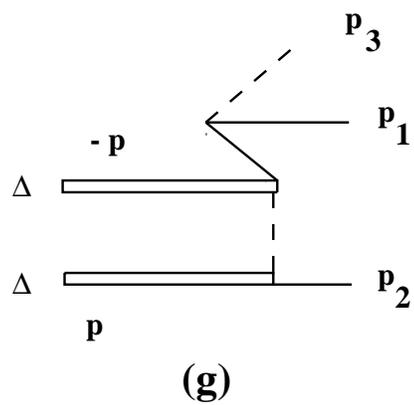
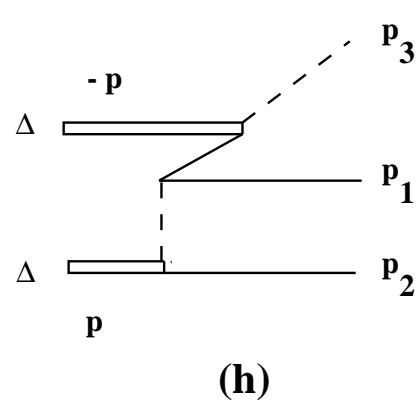
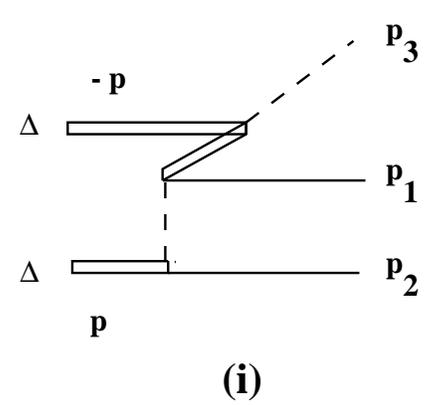
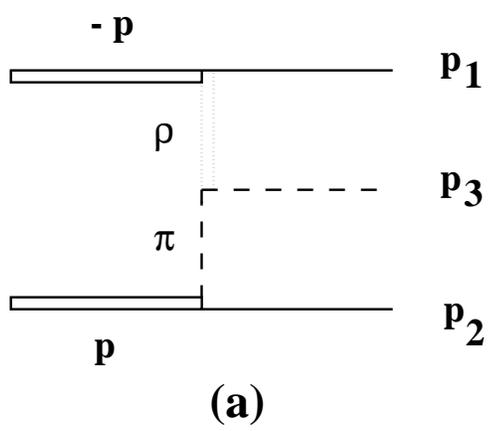
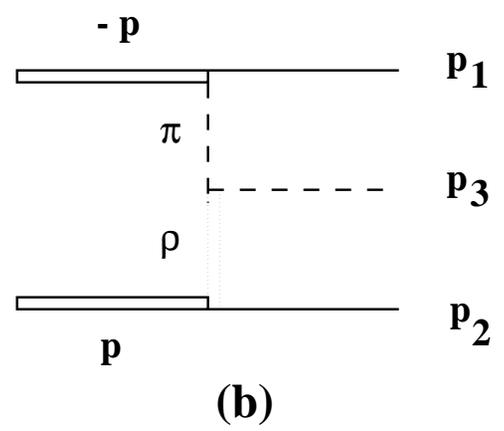
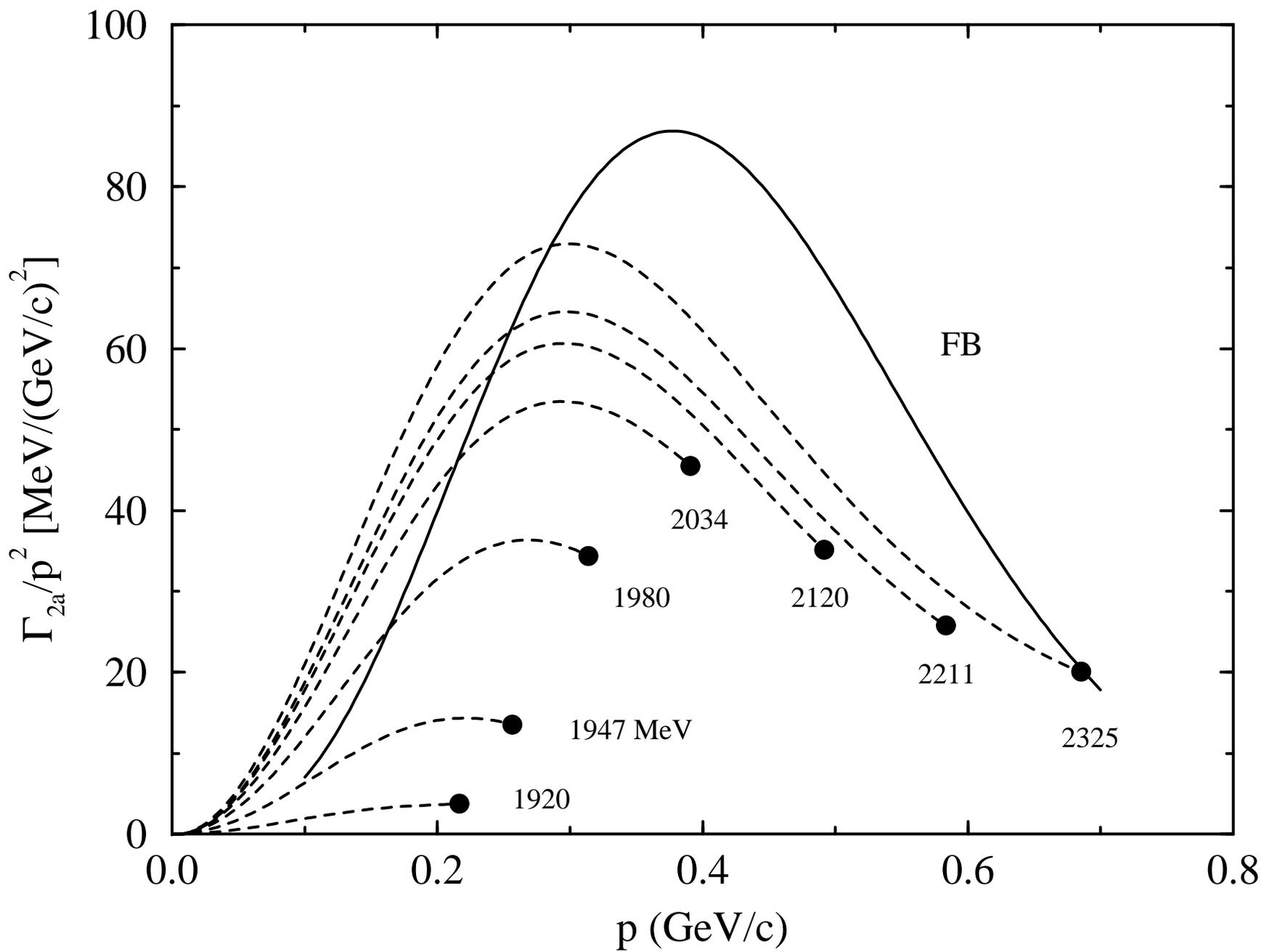
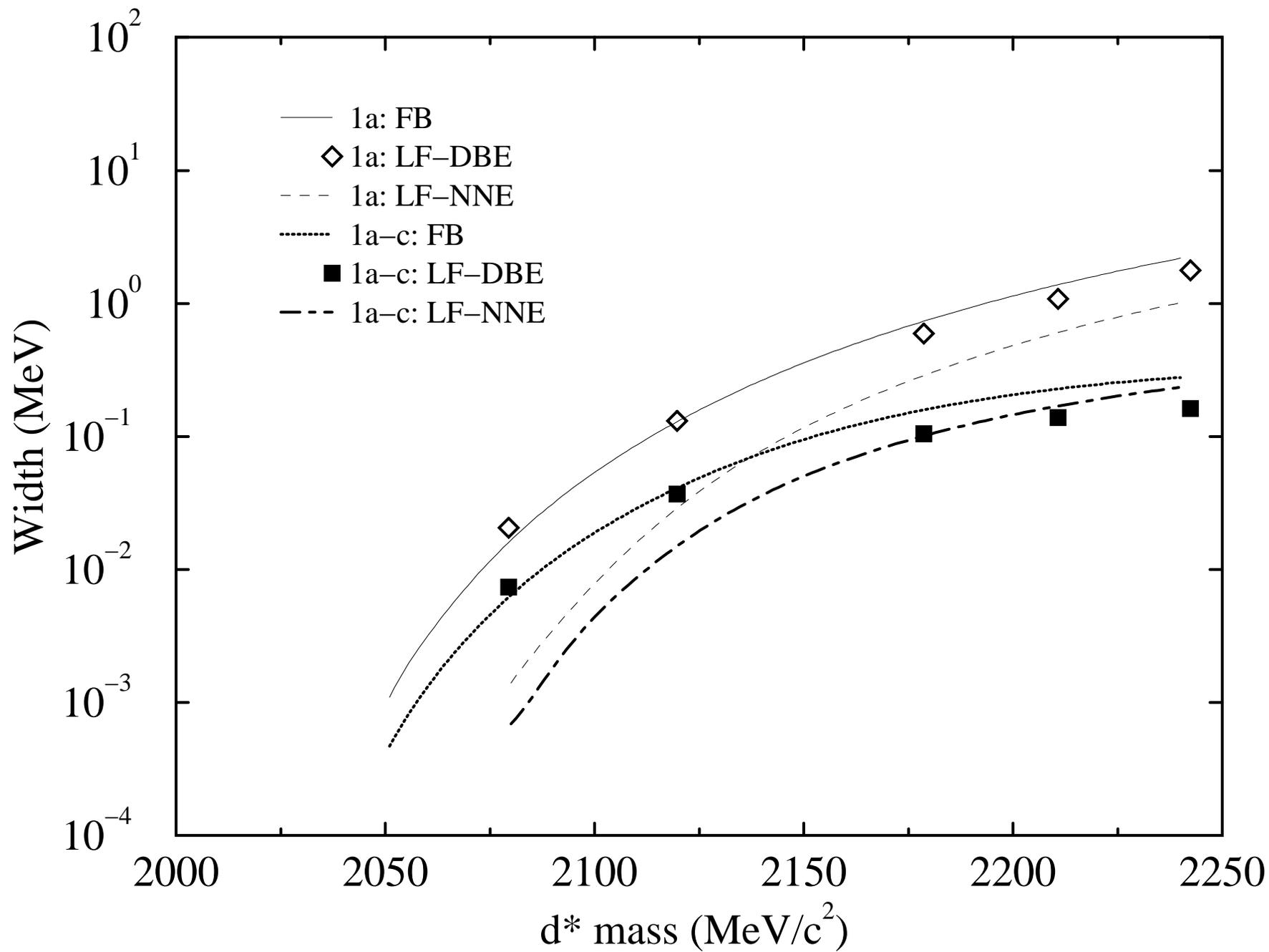
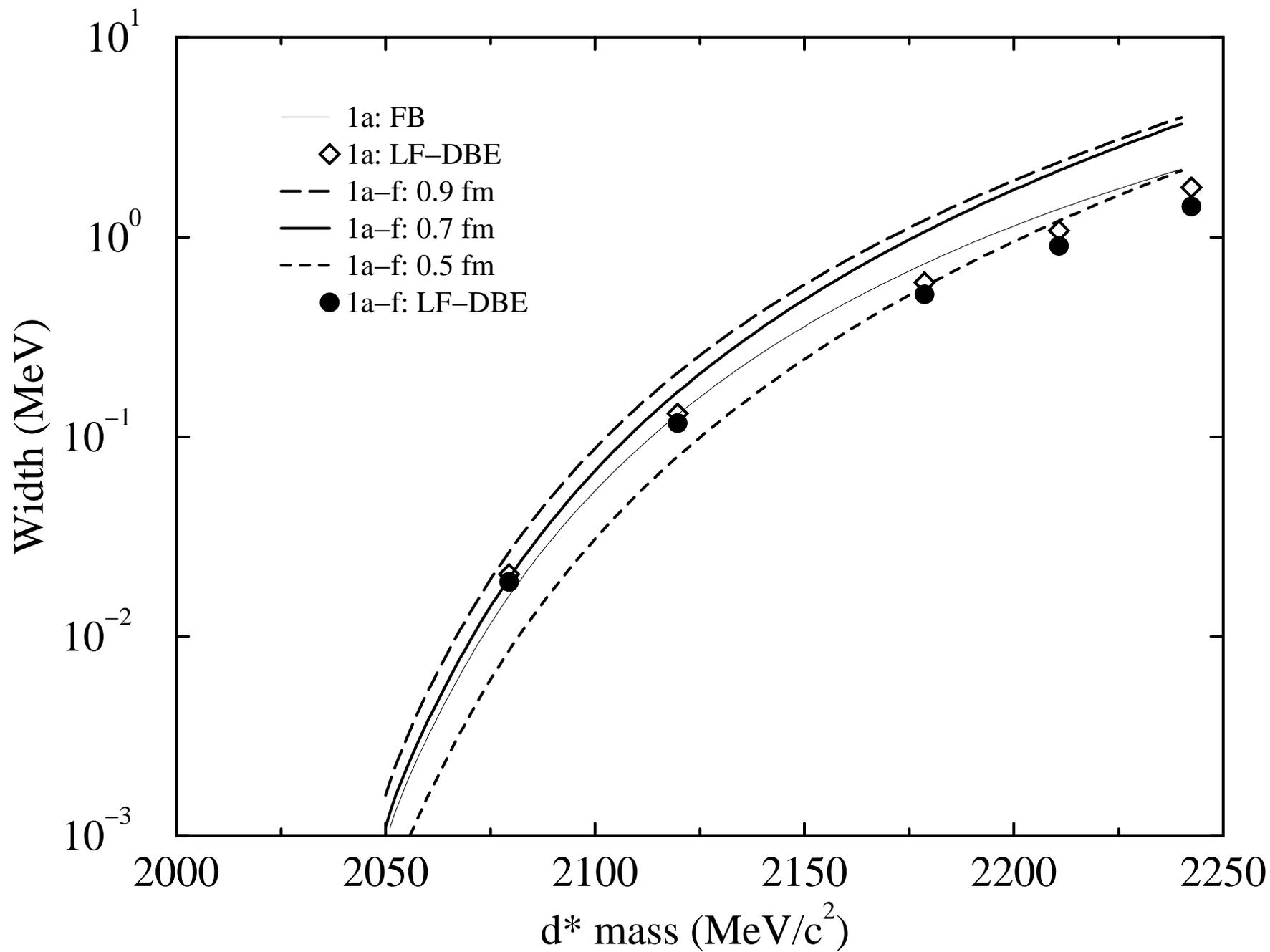
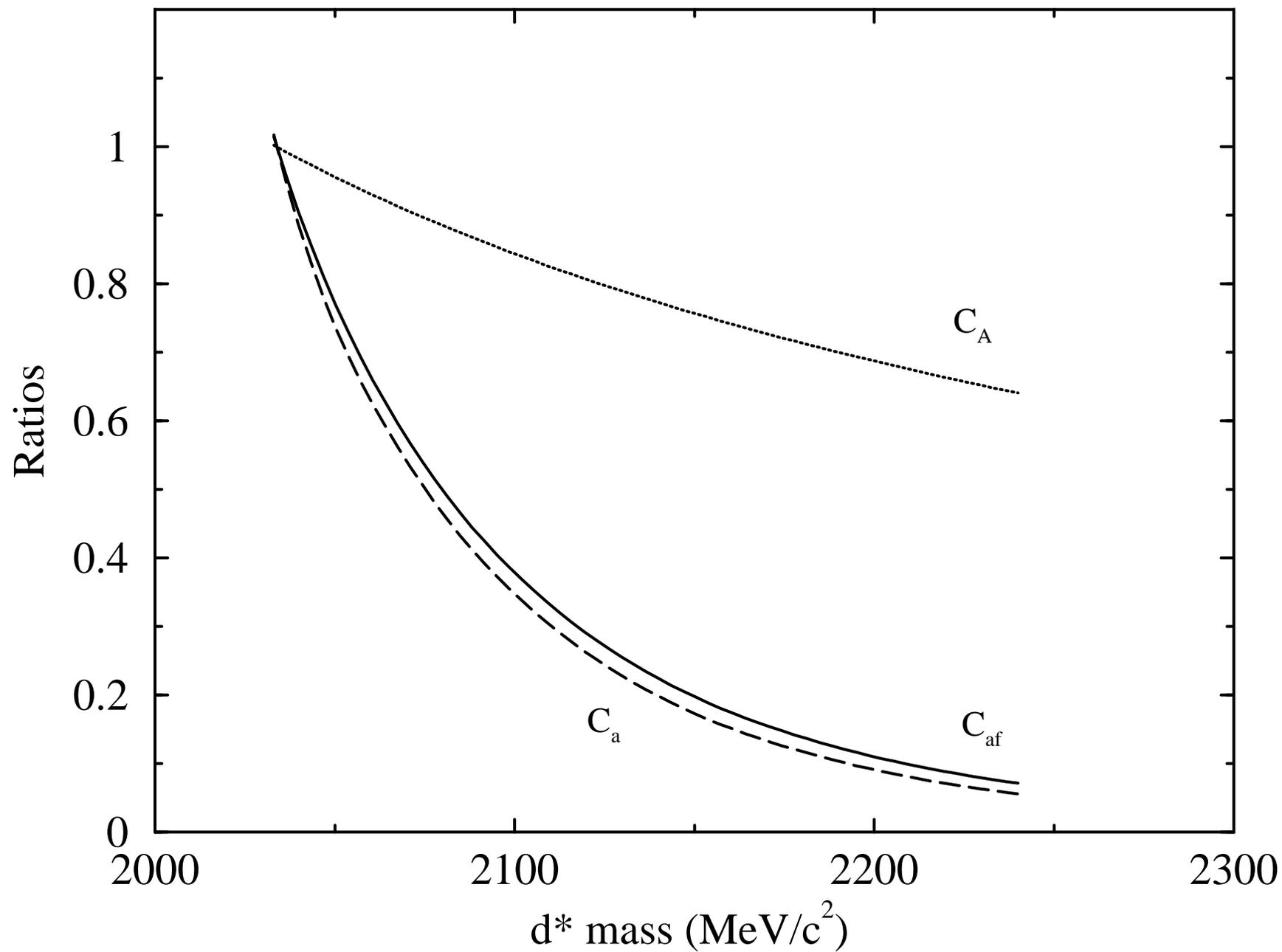
\end{document}